\documentclass[11pt,a4paper]{article}
\usepackage{dsfont}
\usepackage{slashed}
\usepackage{setspace}
\pdfoutput=1
\usepackage[utf8]{inputenc} 
\usepackage{a4wide}
\usepackage{cite}
\usepackage{xcolor}
\usepackage{amssymb}
\usepackage{amsfonts}
\usepackage{graphicx}
\usepackage{hyperref}
\usepackage{mathbbol}
\usepackage{amssymb}
\usepackage{booktabs}
\usepackage{multirow}
\usepackage{appendix}
\DeclareSymbolFontAlphabet{\amsmathbb}{AMSb}
\usepackage{amsfonts}
\usepackage{amssymb}
\usepackage[centertags]{amsmath}

 \def\1{\mathbf{1}}
 \def\3{\mathbf{3}}
 \def\2{\mathbf{2}}

\def\ltap{\ \raisebox{-.4ex}{\rlap{$\sim$}} \raisebox{.4ex}{$<$}\ }
\def\gtap{\ \raisebox{-.4ex}{\rlap{$\sim$}} \raisebox{.4ex}{$>$}\ }

\allowdisplaybreaks[1]

\def\3{\mathbf{3}}

\usepackage{a4wide}
\usepackage{colordvi}

\newcommand{\bec}{\begin{cases}}
\newcommand{\eec}{\end{cases}}
\newcommand{\beq}{\begin{equation*}}
\newcommand{\eeq}{\end{equation*}}
\newcommand{\be}{\begin{equation}}
\newcommand{\ee}{\end{equation}}
\newcommand{\ba}{\begin{eqnarray}}
\newcommand{\ea}{\end{eqnarray}}

\begin{document}
\begin{titlepage}

${}$\vskip 3cm
\vspace*{-15mm}
\begin{flushright}
IPPP/20/39\\
SISSA 21/2020/FISI\\
IPMU20-0095
\end{flushright}
\vspace*{0.7cm}

\vskip 9mm
\begin{center}
{\bf\Large 
Flavoured Resonant Leptogenesis at Sub-TeV Scales
} \\[8mm]
A. Granelli$^{~a}$, K. Moffat$^{~b}$ and 
S.~T.~Petcov$^{~a,c,}$\footnote{Also at:
Institute of Nuclear Research and Nuclear Energy,
Bulgarian Academy of Sciences, 1784 Sofia, Bulgaria.} \\
\vspace{8mm}
$^{a}$\,{\it SISSA/INFN, Via Bonomea 265, 34136 Trieste, Italy.} \\
\vspace{2mm}
$^{b}$\,{\it Institute for Particle Physics Phenomenology, Department of
Physics, Durham University, South Road, Durham DH1 3LE, United Kingdom.}\\
\vspace{2mm}
$^{c}$\,{\it Kavli IPMU (WPI), University of Tokyo, 5-1-5 Kashiwanoha, 277-8583 Kashiwa, Japan.}
\end{center}
\vspace{8mm}

\begin{abstract}
We consider sub-TeV scale flavoured resonant leptogenesis within the 
minimal type-I seesaw scenario with two right-handed singlet neutrinos 
$N_{1,2}$ forming a pseudo-Dirac pair, concentrating on the case of masses 
of the pseudo-Dirac pair having values $M_{1,2} \lesssim 100$ GeV.
The case when the CP violating asymmetries in the individual lepton 
charges $L_l$, $l=e,\mu,\tau$, and in the total lepton charge $L$ 
of the Universe are generated in $1 \leftrightarrow 2$ decay 
processes is investigated.
We show that successful leptogenesis is possible for $M_{1,2}$ 
lying in the interval $M_{1,2} = (0.3 - 100)$ GeV. 
Our results show also, in particular, that for 
vanishing initial $N_{1,2}$ abundance, 
flavour effects  can play an important role in the generation 
of the baryon asymmetry, leading to an enhancement of the asymmetry by a factor 
up to $\sim 300$ with respect to the ``unflavoured'' leptogenesis scenario.
\end{abstract}

\end{titlepage}
\setcounter{footnote}{0}
\setcounter{page}{2}

\section{Introduction}
\label{sec:intro}

Understanding the origin of the excess of matter over antimatter - the matter-antimatter or baryon asymmetry - in the 
Universe remains one of the fundamental problems in particle physics and cosmology. 
The asymmetry can be parametrised by the baryon-to-photon ratio, $\eta_B$, which is defined as
\be
\eta_B \equiv \frac{n_B-n_{\bar{B}}}{n_\gamma},
\ee
%
where $n_B$, $n_{\bar{B}}$ and $n_\gamma$ are the number densities of baryons, anti-baryons and photons, respectively.
The value of $\eta_B$ can be determined using the data on the Cosmic Microwave Background (CMB) radiation \cite{Ade:2015xua}:
\be
{\eta_{B}} _{\text{CMB}} = \left(6.02-6.18\right)\times 10^{-10},
~~95\%~{\rm C.L.}
\ee
%

 A very attractive mechanism of generation of the baryon asymmetry 
is leptogenesis associated with the type-I seesaw scenario of neutrino 
mass generation \cite{Fukugita:1986hr,Kuzmin:1985mm,Minkowski:1977sc,Yanagida:1979as,GellMann:1980vs,Glashow:1979nm,Mohapatra:1979ia}:
it links the existence and smallness of neutrino masses to the existence 
of the baryon asymmetry. Integral to this mechanism are the RH 
neutrinos $\nu_{lR}$ (RH neutrino fields $\nu_{lR}(x)$).
They can be added to the Standard Model (SM) as 
$SU(2)_L\times U(1)_{Y_W}$ singlets without modifying any of the 
fundamental features of the SM. 
The minimally extended SM with two RH neutrinos 
is the minimal scheme in which leptogenesis 
can be realised. The RH neutrinos are assumed to possess a 
Majorana mass term as well as
Yukawa type coupling ${\cal L}_{\rm Y}(x)$ with the Standard Model lepton and Higgs doublets, $\psi_{lL}(x)$ and $\Phi(x)$, respectively. 
In the basis in which the Majorana mass matrix of RH neutrinos and 
the charged lepton mass matrix are diagonal, ${\cal L}_{\rm Y}(x)$ and 
the Majorana mass term have the form:
\be
\label{Ynu}
{\cal L}_{\rm Y,M}(x) =
-\,\left (Y_{li} \overline{\psi_{l L}}(x)\,i\tau_2\,\Phi^*(x)\,N_{iR}(x)
+ \hbox{h.c.} \right )
-\,\frac{1}{2}\,M_{i}\,\overline{N_i}(x)\, N_i(x)\,,
\ee
%
\noindent
where $Y_{li}$ is the matrix of neutrino Yukawa couplings (in the chosen basis) and $N_i$ ($N_i(x)$) is the heavy Majorana neutrino
\footnote{Within the present study the term ``heavy Majorana neutrinos" should be understood to mean Majorana neutrinos 
with masses larger than 100 MeV.
}
(field) possessing a mass $M_i > 0$. 
 
 In what follows we will consider the ``freeze-out" and ``freeze-in'' 
flavoured leptogenesis scenarios in which the Yukawa couplings in 
Eq. \eqref{Ynu} are not CP conserving and the different rates of the decays 
of the Majorana neutrinos $N_j$, $N_j \leftrightarrow l^{+} + \Phi^{(-)}$, 
$N_j \leftrightarrow l^{-} + \Phi^{(+)}$, and of the Higgs boson, 
$ \Phi^{(-)}\rightarrow l^{-} + N_j$, $ \Phi^{(+)}\rightarrow l^{+} + N_j$,
generate CP violating (CPV) asymmetries in the individual lepton 
charges $L_l$, and in the total lepton charge $L$, of the Universe. 
These lepton asymmetries are converted into a baryon asymmetry of the 
Universe (BAU) by $(B-L)$ conserving, but $(B+L)$ violating,
sphaleron processes which exist in the SM and are effective at 
temperatures $T \sim (132 -  10^{12})$ GeV.

The scale and spectrum of masses of the Majorana neutrinos $N_j$ determine the scale of leptogenesis.
In GUT scale leptogenesis $N_j$ have masses a few to several orders smaller than the scale of unification of the electroweak and
strong interactions, $M_{GUT}\cong 2\times 10^{16}$ GeV. 
If the heavy neutrinos $N_j$ have hierarchical spectrum, 
$M_1 \ll M_2 \ll M_3$, the observed baryon asymmetry can be reproduced 
provided the mass of the lightest one satisfies $M_1 \gtap 10^{9}$ GeV 
\cite{Davidson:2002qv}. Moreover, quantitative studies 
in which flavour effects in leptogenesis 
\cite{Nardi:2006fx,Abada:2006fw,Abada:2006ea} (see also                     
\cite{Barbieri:1999ma,Nielsen:2002pc,Endoh:2003mz}) 
were taken into account 
have shown that the CP violation necessary for the generation 
of the observed baryon asymmetry can be provided exclusively by 
the Dirac and/or  Majorana phases in the Pontecorvo, 
Maki, Nakagawa, Sakata (PMNS) neutrino (lepton) mixing matrix $U_{\text{PMNS}}$  
\cite{Pascoli:2006ie,Pascoli:2006ci,Blanchet:2006be,Branco:2006ce,Anisimov:2007mw,Molinaro:2008rg,Molinaro:2008cw,Dolan:2018qpy}. 
More recent analyses revealed \cite{Moffat:2018wke,Moffat:2018smo} 
that in the case of heavy Majorana neutrino mass spectrum with mild hierarchy, 
$M_2 \sim 3M_1$, $M_3\sim 3M_2$,
i) successful leptogenesis can take place for $M_1 \gtap 10^{6}$ GeV, 
and that ii) also in this case the required CP violation can be provided 
exclusively by the Dirac or Majorana CPV phases of the neutrino mixing matrix.
In \cite{Brivio:2019hrj} this was confirmed to be the case as well in the so-called ``Neutrino Option" seesaw scenario 
\cite{Brivio:2017dfq} in which the mass term in the Higgs potential, responsible for the electroweak symmetry breaking 
in the Standard Theory, is generated at one loop level by the neutrino Yukawa coupling in Eq. (\ref{Ynu}).  
In the ``Neutrino Option" scenario with two Majorana neutrinos $N_{1,2}$, 
successful leptogenesis was shown to be possible only in the so-called ``resonant regime" 
\cite{Pilaftsis:2003gt,Dev:2014laa,Dev:2014wsa}, with $N_{1,2}$ forming 
a pseudo-Dirac pair \cite{Wolfenstein:1981kw,Petcov:1982ya} with masses $M\equiv (M_1 + M_2)/2 \sim (1-8)\times 10^6$ GeV and 
splitting between them, which is of the order of the $N_{1,2}$ decay widths $\Gamma_{1,2}$: 
$\Delta M/\Gamma_{1,2}\sim 1$, $\Delta M/M \equiv (M_2 - M_1)/M \sim 10^{-8}$.

  One attractive feature of Resonant Leptogenesis (RL) is that 
the baryon asymmetry can be produced at relatively low scales, e.g., 
at the TeV scale. Studies have shown that it is possible to have successful 
RL at scales exceeding approximately 100 GeV
(see, e.g., \cite{Bambhaniya:2016rbb} and references quoted therein) 
or even at smaller scales if thermal effects are taken into account 
leading to the possibility of CPV Higgs decays into $N_{1,2}$ plus a 
lepton \cite{1606.00017}.
Scenarios with low scale RL typically lead to predictions that potentially 
can be tested at colliders (LHC or future planned) and/or at 
low-energy experiments (see, e.g., \cite{Bambhaniya:2016rbb,1606.00017}).

In the present article we consider sub-TeV scale flavoured RL 
within the minimal type-I seesaw scenario with two (RH) singlet 
neutrinos $N_{1,2}$ forming a pseudo-Dirac pair. 
We concentrate on the case when the masses of 
the pseudo-Dirac pair have values $M_{1,2} \ltap 100$ GeV 
and consider scenarios in which the baryon asymmetry is generated 
in CP violating Higgs and $N_{1,2}$ decays.  
We do not consider in this study the ``freeze-in'' 
scenario \cite{Akhmedov:1998qx,Asaka:2005pn} 
in which the BAU in leptogenesis is generated
via $N_1 \leftrightarrow N_2$ oscillations during the epoch when 
$N_{1,2}$ are out of equilibrium, which has been extensively studied 
(see, e.g., \cite{hepph0605047,1112.5565,canetti2013dark,Shuve:2014zua,1508.03676,1606.06690,Hernandez:2016kel,1704.02692,1703.06087,Drewes:2017zyw} 
and references quoted therein) 
\footnote{
In the ``freeze-in'' $N_1-N_2$ oscillation scenario 
\cite{Akhmedov:1998qx}
the generation of the baryon asymmetry proceeds, 
as was shown in \cite{Asaka:2005pn}, 
principally via a lepton number conserving (LNC) terms  
involving fourth power of the  neutrino Yukawa couplings.
In the scenario considered by us 
the baryon asymmetry is generated predominantly by 
lepton number violating (LNV) terms (involving also fourth power of 
the neutrino Yukawa couplings).
}.

Our study differs from the study performed in 
\cite{1606.00017}, in which the Higgs decay scenario has first been 
considered, in two aspects: i) we use the improvements in accounting
for thermal effects in the processes of interest, which 
have appeared in the literature since the basic study \cite{hepph0310123} 
the results of which were employed in \cite{1606.00017}, 
and ii) we take into account the flavour effects, 
which were not accounted for in \cite{1606.00017}. 
We find, in particular, that flavour effects 
can play very important role in RL of interest especially 
in the case of zero initial abundance of the heavy 
Majorana neutrinos. 

Our work differs also from the studies reported in 
\cite{1705.00016,1710.03744}. In \cite{1705.00016} the authors 
investigated the generation of the baryon asymmetry by both 
the Higgs decay and the ``freeze-in'' oscillation mechanisms 
without separating the contributions of each of the two mechanisms. 
The aim of our work is, in particular, to identify just the 
Higgs decay contribution to the generation of BAU.
In \cite{1705.00016} flavour effects are taken into account, 
but the enhancement of the range of the Higgs decay contribution 
due to thermal effects 
(the collinear emissions of soft gauge bosons (see further)), 
is not. In our study the latter effects are accounted for. 
We have considered also two different types of initial conditions 
for the two heavy Majorana neutrinos: thermal initial abundance (TIA) 
and vanishing (zero) initial abundance (VIA). In \cite{1705.00016} 
only the case of VIA has been investigated.

In \cite{1710.03744} the authors also 
studied the generation of the baryon asymmetry by both 
the Higgs decay and the ``freeze-in'' oscillation mechanisms 
including the flavour effects, but without separating the contributions 
of each of the two mechanisms. Only the VIA case has been considered.
In addition the masses of $N_{1,2}$ were constrained to lie 
in the interval $M_{1,2} = 5 - 50$ GeV. In our study
the minimal $M_{1,2}$ is determined by the requirement of reproducing 
the observed value of BAU varying all other parameters in the problem. 
We find, in particular, that in the VIA case 
the minimal $M_{1,2}$ is by more than a factor of 10 smaller than the value
assumed in \cite{1710.03744}. 

In the relatively recent article  \cite{Klaric:2020lov} 
the authors presented a unified framework of generation of the baryon asymmetry at the sub-TeV scale via the neutrino oscillations 
(``freeze-in'') and the decay RL mechanisms within the scenario with two Majorana 
neutrinos $N_{1,2}$. They have shown that 
i) the observed baryon asymmetry can be generated for 
all experimentally allowed values of the Majorana neutrino masses 
$M_{1,2} \gtrsim 100$ MeV and up to $M_{1,2} \sim 1$ TeV, and that 
ii) leptogenesis is effective in a broad range of 
the relevant parameters, including mass splitting between 
the two Majorana neutrinos 
as big as $(M_2-M_1)/(0.5(M_2 +M_1))\sim 0.1$, as well as 
couplings of $N_{1,2}$ in the weak charged lepton current 
large enough to be accessible to planned intensity experiments 
or future colliders. The results are presented in  
\cite{Klaric:2020lov} without separation of the contributions 
of the ``freeze-in'' neutrino oscillation mechanism and the decay 
mechanism in the VIA case. In the TIA case the separation of interest 
is made only in Fig. 2 and we find similar lower bound on $M_{1,2}$ 
as reported in \cite{Klaric:2020lov}, with the region of the parameter 
space of successful RL reported in \cite{Klaric:2020lov} being somewhat 
larger than that we found in our work. As we have already emphasised, 
we have concentrated in our work on the decay mechanism of 
baryon asymmetry generation. We wanted to identify the parameter 
space in which the observed value of BAU could be generated 
via the decay mechanism.  
Thus, effectively we explore part of the parameter 
space explored in \cite{Klaric:2020lov}. 

We should add finally that in 
\cite{1705.00016,1710.03744,Klaric:2020lov} 
the authors use the density matrix formalism for the calculation of the 
baryon asymmetry, while we use a system of Boltzmann equations 
with a proper source term for the treatment of the resonance effects 
in leptogenesis based on the decay mechanism and take into account the 
flavour and thermal effects in the decay mechanism. 
In the strong wash-out regime, in which the results 
of our study are obtained, we expect the results based on the two approaches 
and obtained within the decay mechanism to be largely compatible, 
with differences that should not exceed a factor of $\sim (2-3)$ 
in the value of the generated baryon asymmetry. 
For the reasons explained earlier it is impossible at present to 
make a comparison between our results and those obtained in 
\cite{1705.00016,1710.03744,Klaric:2020lov} in the VIA case; 
in the TIA case a comparison is possible to a certain degree with 
the results reported in \cite{Klaric:2020lov} (see Section 3.3).

 The paper is organised as follows. In Section \ref{sec:nu}, 
we summarise the basics of the type I seesaw scenario 
and the conventions we will employ throughout.
 In Section \ref{sec:thermeff} we introduce the equations relevant 
for RL at scales $T \lesssim 10^3$ GeV of interest. 
Then we proceed to show results in two possible scenarios 
by which the BAU can be produced. They correspond to two different 
``initial conditions'', i.e.,  $N_{1,2}$ initial abundances: 
i) $N_{1,2}$ thermal initial abundance (TIA), and 
ii) $N_{1,2}$  vanishing (zero) initial abundance (VIA).
We conclude in Section \ref{sec:concs} with a brief
summary of our results.

\section{Seesaw, Neutrino Masses and Mixing }
\label{sec:nu}

 In the present Section we set the notations and 
review some of the elements of the 
seesaw theory that will be used in our further analysis 
(see, e.g., \cite{Ibarra:2010xw}).

 In the basis in which the charged lepton Yukawa couplings 
and mass matrix are diagonal but the Majorana mass term of the 
RH neutrinos $\nu_{lR}$ is not, the Lagrangian ${\cal L}_{\rm Y,M}(x)$ 
has the form:
\be
\label{Ynu2}
{\cal L}_{\rm Y,M}(x) =
-\,\tilde{Y}_{ll'} \overline{\psi_{l L}}(x)\,i\tau_2\,\Phi^*(x)\,
\nu_{l'R}(x)
-\,\frac{1}{2}\,\overline{\nu^C_{lL}}(x) \,(M_{N})_{ll'}\, \nu_{l'R}(x)\,
+ \hbox{h.c.}\,,
\ee
%
\noindent
where $\tilde{Y}$ is the matrix of neutrino Yukawa couplings in 
the considered basis, 
$(\psi_{lL}(x))^T = (\nu^T_{lL}(x)~~l^T_{L}(x))$, $l=e,\mu,\tau$,
$\nu_{lL}(x)$ and $l_{L}(x)$ being the left-handed (LH) flavour neutrino 
and charged lepton fields, 
$(\Phi(x))^T = (\Phi^{(+)}(x)~\Phi^{(0)}(x))$,
$\nu^C_{lL}(x) = C\,(\overline{\nu_{lR}}(x))^T$,
$C$ being the charge conjugation matrix, and 
$M_N$ is the Majorana mass matrix of  $\nu_{lR}(x)$, 
$M^T_N = M_N$. When the electroweak
symmetry is broken spontaneously,
the neutrino Yukawa coupling in Eq. (\ref{Ynu2})
generates a Dirac mass term,
$(M_{D})_{ll'}\,\overline{\nu_{l L}}(x)\, \nu_{l'R}(x) + \hbox{h.c.}$,
with $M_{D} = (v/\sqrt{2})\tilde{Y}$, $v = 246$ GeV
being the Higgs doublet vacuum expectation value (VEV),
and the neutrino mass Lagrangian takes the form:
\begin{equation}
\begin{split}
\mathcal{L}^m_\nu&=-\,\overline{\nu_{lL}}(M_D)_{ll'}\nu_{l'R} 
-\,\frac{1}{2}\overline{\nu^c_{lL}}(M_N)_{ll'}\,\nu_{l'R}+h.c.=\\
&= -\,\frac{1}{2}\begin{pmatrix}
\overline{\nu_{\alpha L}}&\overline{\nu^c_{\kappa L}}
\end{pmatrix}
\begin{pmatrix}\mathbb{O}_{\alpha\beta}&(M_D)_{\alpha \rho}
\\(M_D^T)_{\kappa \beta}&(M_N)_{\kappa \rho}
\end{pmatrix}
\begin{pmatrix}
\nu^c_{\beta R}\\\nu_{\rho R}
\end{pmatrix}+h.c.\,,
\end{split}
\label{Lnum}
\end{equation}
%
where $\nu^c_{\beta R}\equiv C (\overline{\nu_{\beta L}})^T$ 
and $\alpha,\beta = e,\mu,\tau$; in the case of three 
right-handed neutrinos we can choose
\footnote{The labelling of the right-handed neutrinos 
does not need to coincide with the labelling we use 
for the left-handed neutrinos; we can choose the indices 
$\kappa$ and $\rho$ to take, e.g., the values 
$\kappa,\rho=\tilde{e},\tilde{\mu},\tilde{\tau},...,\tilde{\sigma}$.
}
$\kappa,\rho=e,\mu,\tau$.
The two matrices $M_D$ and $M_N$ are complex, in general. 

The diagonalisation of the mass term under the condition 
that $M_D$ is much smaller than $M_N$~
\footnote{More precisely, the condition requires that  
the elements of $M_{D}$ are much smaller than the eigenvalues $M_k$ of 
$M_N$.} 
leads to the well-known effective Majorana mass
(term) for the LH flavour neutrinos 
\cite{Minkowski:1977sc,Yanagida:1979as,GellMann:1980vs,Glashow:1979nm,Mohapatra:1979ia}:
\be
m_\nu \cong -\, M_{D}\,M^{-1}_{N}\,(M_D)^{T} 
= U\, \hat{m}_\nu\, U^T\,,  
\label{seesawnuMajM}
\ee
%
where $\hat{m}_\nu = {\rm diag}(m_1,m_2,m_3)$, $m_{1,2,3}$ 
being the masses of the light Majorana neutrinos $\nu_{1,2,3}$, 
$m_i \ltap 0.5$ eV, 
and $U$ is a $3\times 3$ unitary matrix. 
The flavour neutrino fields are related to the fields 
of light and heavy neutrinos $\nu_i(x)$ and $N_j(x)$
with definite mass $m_i$ and $M_j$, $m_i \ll M_j$, via 
\be
\nu_{lL}(x) = \sum_j (1 + \eta)U_{lj}\nu_{jL}(x) + (RV)_{lj} N_{jL}(x)\,.
\label{nulnuj}
\ee
%
Here $\nu_{jL}(x)$ and $N_{jL}(x)$ are the left-handed components 
of  $\nu_i(x)$ and $N_j(x)$,
$R \cong M_DM^{-1}_N$, 
$\eta = -\,\dfrac{1}{2}RR^\dagger =  -\,\dfrac{1}{2}(RV)(RV)^\dagger$ and 
$V$ is a unitary matrix 
which (to leading approximation in $M_D/M_N$) diagonalises the 
Majorana mass matrix of the RH neutrinos $M_N$.
The heavy neutrinos $N_j$ are mass-eigenstates of $M_N$.
The constants $(RV)_{lj}$ represent the weak charged and neutral 
current couplings of the heavy Majorana neutrinos.
There exist stringent upper limits on the elements of $\eta$, and thus 
on the elements of $RV$, from electroweak data and data 
on flavour observables \cite{Fernandez-Martinez:2015hxa,Blennow:2016jkn}.  
For $M_j \gtap 500$ MeV, depending on the 
element of $\eta$, they are in the range of $10^{-3} - 10^{-4}$ 
at $2\sigma$ C.L. For $M_j$ larger than the electroweak scale, 
the constraint on $\eta_{e\mu} = \eta_{\mu e}$ is even stronger: 
$|\eta_{e\mu}| < 1.2\times 10^{-5}$. 
 
The PMNS matrix (in the diagonal charged lepton mass basis 
employed by us) has the form:
\begin{equation}
U_{\text{PMNS}} = (1+\eta)\,U\,.
\end{equation}
%
Thus, the matrix $\eta$ parametrises the departure from unitarity of 
the PMNS matrix. Given the existing limits on the elements of 
$\eta$, we have to a very good approximation: 
$U_{\text{PMNS}} \cong U$.
We will use in what follows the standard parametrisation 
of the PMNS matrix U \cite{Tanabashi:2018oca}:
\begin{equation}
\label{PMNS}
U = \begin{pmatrix}
c_{12}c_{13}&s_{12}c_{13}&s_{13}\text{e}^{-i\delta}\\
-s_{12}c_{23}-c_{12}s_{23}s_{13}\text{e}^{i\delta}&c_{12}c_{23}-s_{12}s_{23}s_{13}\text{e}^{i\delta}&s_{23}c_{13}\\
s_{12}s_{23}-c_{12}c_{23}s_{13}\text{e}^{i\delta}&-c_{12}s_{23}-s_{12}c_{23}s_{13}\text{e}^{i\delta}&c_{23}c_{13}
\end{pmatrix}\times
\begin{pmatrix}
1&0&0\\
0&\text{e}^{\frac{i\alpha_{21}}{2}}&0\\
0&0&\text{e}^{\frac{i\alpha_{31}}{2}}
\end{pmatrix},
\end{equation}
%
where $c_{ij} \equiv \cos\theta_{ij}$, $s_{ij} \equiv \sin\theta_{ij}$, 
$\delta$ is the Dirac CP violation (CPV) phase, 
and $\alpha_{21}$ and $\alpha_{31}$ are the two Majorana 
CPV phases \cite{Bilenky:1980cx}.

As is well-known, the mass spectrum of neutrinos $\nu_{1,2,3}$ 
can be with normal ordering (NO), $m_1 < m_2 < m_3$, 
or with inverted ordering (IO), $m_3 < m_1 < m_2$ 
(see, e.g., \cite{Tanabashi:2018oca}). In what follows we will 
concentrate on the case of NO neutrino mass spectrum.

 As we have already indicated, we will consider the type-I seesaw 
scenario with only two ``heavy" (singlet) Majorana neutrinos $N_{1,2}$. 
This is the minimal scenario compatible with the oscillation data 
\cite{Tanabashi:2018oca}. In this case the lightest of the three neutrinos 
$\nu_{1,2,3}$ is massless at tree and one-loop level, 
$m_1 \cong 0$ (NO spectrum) and we have: 
$m_2 = \sqrt{\Delta m^2_{21}}$, $m_3 = \sqrt{\Delta m^2_{31}}$,
where $\Delta m^2_{ij} \equiv m^2_i - m^2_j$.
The neutrino mass spectrum is normal hierarchical 
(NH): $m_1 \ll m_2 \ll m_3$. Of the two Majorana phases, 
 $\alpha_{21}$ and $\alpha_{31}$,
only the phase difference $\alpha_{21} - \alpha_{31} \equiv \alpha_{23}$, is physical \footnote{We will call $\alpha_{23}$ ``Majorana phase'' in what follows.}. 
Technically, we will use the formalism employed for the presence of 
three heavy Majorana neutrinos $N_{1,2,3}$ in which, however, 
$m_1 = 0$ and $N_3$ is decoupled.

In our numerical analyses we will use the values 
of the three neutrino mixing angles $\theta_{12}$,  $\theta_{23}$ 
and  $\theta_{13}$, and the two neutrino mass squared differences 
obtained in the global neutrino oscillation data analysis performed 
in \cite{Esteban_2020} and quoted in Table \ref{tab:PMNSparams}.
\begin{table}[t]
\begin{tabular}{l|ll|ll|}
\cline{2-5}
                       & \multicolumn{2}{c|}{\textbf{NO}}                       & \multicolumn{2}{c|}{\textbf{IO}}                       \\ \cline{2-5} 
                       & \multicolumn{1}{|c}{Best Fit} & \multicolumn{1}{c|}{$3\sigma$} & \multicolumn{1}{|c}{Best Fit} &\multicolumn{1}{c|}{$3\sigma$} \\ \hline
\multicolumn{1}{|r|}{$\theta_{12}\;(^\circ)$}				&	\multicolumn{1}{c}{$33.44$}	&	\multicolumn{1}{c|}{$[31.27,35.86]$}	&	\multicolumn{1}{|c}{$33.45$}	&	\multicolumn{1}{c|}{$[31.27,35.87]$}	\\
\multicolumn{1}{|r|}{$\theta_{13}\;(^\circ)$}				&	\multicolumn{1}{c}{$8.57$}	&	\multicolumn{1}{c|}{$[8.20,8.93]$}	&	\multicolumn{1}{|c}{$8.60$}	&	\multicolumn{1}{c|}{$[8.24,8.96]$}	\\
\multicolumn{1}{|r|}{$\theta_{23}\;(^\circ)$}				&	\multicolumn{1}{c}{$49.2$}	&	\multicolumn{1}{c|}{$[40.1,51.7]$}	&	\multicolumn{1}{|c}{$49.3$}	&	\multicolumn{1}{c|}{$[40.3,51.8]$}	\\
\multicolumn{1}{|r|}{$\delta\;(^\circ)$}					&	\multicolumn{1}{c}{$197$}	&	\multicolumn{1}{c|}{$[120,369]$}	&	\multicolumn{1}{|c}{$282$}	&	\multicolumn{1}{c|}{$[193,352]$}	\\
\multicolumn{1}{|r|}{$\Delta m_{21}^2\;(\cdot 10^{-5}\text{eV}^2)$}		&	\multicolumn{1}{c}{$7.42$}	&	\multicolumn{1}{c|}{$[6.82,8.04]$}	&	\multicolumn{1}{|c}{$7.42$}	&	\multicolumn{1}{c|}{$[6.82,8.04]$}	\\
\multicolumn{1}{|r|}{$\Delta m_{31(32)}^2\;(\cdot 10^{-3}\text{eV}^2)$} 	&	\multicolumn{1}{c}{$2.517$}	&	\multicolumn{1}{c|}{$[2.435,2.598]$}	&	\multicolumn{1}{|c}{$-2.498$}	&	\multicolumn{1}{c|}{$[-2.581,-2.414]$}	\\
\hline
\end{tabular}
\caption{The best fit values and $3\sigma$ ranges for the parameters of 
the PMNS matrix $U$ and the square mass differences in the Normal Ordering (NO) and Inverted Ordering (IO) cases \cite{Esteban_2020}. Notice that the 
3$\sigma$ range for the Dirac phase $\delta$ is quite large, so we will treat 
it as an unmeasured parameter.}
\label{tab:PMNSparams}
\end{table}
%

Equation (\ref{seesawnuMajM}) 
allows to relate the matrix of the neutrino Yukawa couplings 
$Y$ and the PMNS matrix $U$ \cite{Casas:2001sr}. 
In the diagonal mass basis of  the heavy Majorana neutrinos, 
which is convenient to use in the leptogenesis analyses, 
we have:
\be 
Y = \tilde{Y}\,V^* =  
 i\, \dfrac{\sqrt{2}}{v}\,U\, \sqrt{\hat{m}_\nu}\,O^T\sqrt{\hat{M}}\,,
\label{CI}
\ee
%
where $O$ is a complex orthogonal matrix, $O^T\,O = O\,O^T = I$.
In the case of interest with two active ``heavy" 
Majorana neutrinos $N_{1,2}$ and the formalism employed by us, 
we have $\hat{M} = {\rm diag}(M_1,M_2,M_3)$, 
but with $m_1 = 0$ and the form of the $O$ matrix given 
for the NH spectrum below, the mass $M_3$ of the third decoupled 
heavy Majorana neutrino does not play any role in our analysis:
\begin{equation}
\label{NHO}
O=\begin{pmatrix}
0&\cos \theta&\sin\theta\\
0&-\sin\theta&\cos\theta\\
1&0&0
\end{pmatrix}=\frac{e^{-i\omega}e^{\xi}}{2}\begin{pmatrix}0&1&-i\\0&i&1\\1&0&0\end{pmatrix}+\frac{e^{i\omega}e^{-\xi}}{2}\begin{pmatrix}0&1&i\\0&-i&1\\1&0&0\end{pmatrix},
\end{equation}
%
where $\theta=\omega+i\xi$, $\omega$ and $\xi$ being two real parameters 
\footnote{With $m_1 = 0$, the form of the matrix $O$ in Eq. \eqref{NHO}, as can be easily 
checked, corresponds to the case of decoupled $N_3$.}.
The parameters $\omega$ and $\xi$ 
play important roles in the leptogenesis scenario we are going to consider 
next. For large values of $\xi$, such that $e^\xi\gg e^{-\xi}$ the first 
term of the above expression 
dominates, being enhanced by the exponential. 
The $RV$-matrix too is enhanced by large values of $\xi$ and the sum 
of the square modulus of its entries reads (NH):
\be\label{RV}
\sum_{l, i}|(RV)_{li}|^2\simeq \frac{1}{2 M}(m_2+m_3) e^{2\xi}.
\ee
%

The $O$-matrices in Eqs. \eqref{NHO} has  det$(O) = 1$. 
In the literature, the  factor 
$\varphi = \pm 1$ is sometimes included in the definition of $O$ to allow 
for the both cases det$(O) = \pm 1$. We choose instead to 
work with the matrix in  Eqs. \eqref{NHO} but
extend the range of the Majorana phases $\alpha_{21(31)}$ 
from  $[0, 2\pi]$ to $[0, 4\pi]$, 
which effectively accounts for the case of  det$(O) = -\,1$ 
\cite{Molinaro:2008rg}, 
so that the same full set of $O$ and Yukawa matrices is considered.

%
\section{Flavoured Resonant Leptogenesis at sub-100 GeV scales}
\label{sec:thermeff}
%

 In this study we solve the Boltzmann system of equations for 
the $N_{1,2}$ and BAU abundances taking into account both 
$1\leftrightarrow 2$ decays and inverse decays
and $2\leftrightarrow 2$ scatterings including the thermal effects. 
In the case of three-flavoured RL of interest it has the form 
(see, e.g., \cite{Nardi:2006fx,Blanchet:2011xq,Davidson:2008bu})
\footnote{These equations approximate the results of 
\cite{Dev:2014laa,Dev:2014wsa} for RL. The latter results should agree 
with those of \cite{Garbrecht:2011aw} to within a factor $\sim 2$ 
\cite{Millington} in the nearly degenerate mass regime considered 
in this article.}:
\begin{eqnarray}
\label{BEsRLThN}
    \frac{d N_{N_i}}{dz}             &=& - \left(D_i + S_i^t + S_i^s\right) (N_{N_i}-N_{N_i}^{eq}), \\
\label{BEsRLThL}
    \frac{dN_{\Delta_\alpha}}{dz}   &=& \sum_i \left[- \epsilon_{\alpha\alpha}^{(i)} D_i (N_{N_i}-N_{N_i}^{eq}) - \left(W_i^D + + W_i^t + 
                                                                            \frac{N_{N_i}}{N_{N_i}^{eq}} W_i^s\right) p_{i\alpha} N_{\Delta_\alpha}\right],
\end{eqnarray}
%
where $z \equiv M/T$, $M_{1,2} \cong M$. 
The quantities $N_{N_i}$ and $N_{\Delta_\alpha}$ are respectively the number 
of heavy neutrinos $N_i$ and the value of the asymmetry 
$\Delta_\alpha \equiv\frac{1}{3}B-L_\alpha$, 
$\alpha = e,\,\mu,\,\tau$, in a comoving volume, 
normalised to contain one photon at $z=0$, i.e., $N^{eq}_{N_i}(0)=3/4$. 
This normalisation within the Boltzmann statistics is equivalent 
to using $N_{N_i}^{eq}(z) = \frac{3}{8}z^2 K_2(z)$, 
where $K_n(z)$, $n=1,2,..$, are the modified 
 $n^\text{th}$ Bessel functions of the second kind.
 
 Equations \eqref{BEsRLThN} and \eqref{BEsRLThL} are an excellent 
approximation to the density matrix equations of \cite{Blanchet:2011xq}, 
which reduce to Eqs. \eqref{BEsRLThN} and \eqref{BEsRLThL} for 
$T \lesssim 10^7$ GeV (the case in which all lepton flavours are 
fully decohered). 
We note that they do not include relativistic corrections and also they 
are written under the assumption of kinetic equilibrium.

The term $p_{i\alpha}$ which multiplies the wash-outs, is the projection probability of heavy neutrino mass state $i$ on to flavour state $\alpha$ 
and is given by
\begin{equation}
    p_{i\alpha} = \frac{\left|Y_{\alpha i}\right|^2v^2}{2\tilde{m}_iM_i},
\end{equation}
%
with $\tilde{m}_i\equiv(Y^\dagger Y)_{ii}v^2/2M_i$. 
The projection probabilities $p_{i \alpha}$, $i=1,2$, 
$\alpha = e,\mu,\tau$, are strongly flavour dependent.

 We use the following conversion from the $B - L$ 
number density to the baryon-to-photon ratio:
\begin{equation}
    \eta_B = \eta_{B_e} + \eta_{B_\mu} + \eta_{B_\tau} 
= \frac{28}{79}\frac{1}{27}
\left (N_{\Delta_e} + N_{\Delta_\mu} +  N_{\Delta_\tau}\right)\,,
\label{eq:etaBl}
\end{equation}
%
where $28/79$ is the SM sphaleron conversion coefficient
and the $1/27$ factor comes from the dilution of the baryon asymmetry 
by photons.

%
\subsection{The Decay and Scattering Terms}
%

The terms $D_i$ and $W^D_i$ are due to the 
$ 1 \leftrightarrow 2$ decays and inverse decays, while
\begin{eqnarray}
\label{eq:St}
    S^t_i &=& 4\,\left [S^{(\text{gauge})}_{A_t i} +  S^{(\text{quark})}_{H_t i}\right ]\,,\\
\label{eq:Ss}
    S^s_i &=&  2\,\left [S^{(\text{gauge})}_{A_s i} +  S^{(\text{quark})}_{H_s i}\right ]\,,\\
\label{eq:Wt}
    W^t_i &=& 4\,\left [W^{(\text{gauge})}_{A_t i} + W^{(\text{quark})}_{H_t i}\right ]\,,\\
\label{eq:Ws}
    W^s_i &=& 2\,\left [W^{(\text{gauge})}_{A_s i} + W^{(\text{quark})}_{H_s i}\right ]\,,
\end{eqnarray}
%
account for scattering processes.
The terms $S^{(\text{gauge})}_{A_t i}$
($W^{(\text{gauge})}_{A_t i}$) and 
$S^{(\text{gauge})}_{A_s i}$ 
($W^{(\text{gauge})}_{A_s i}$) 
are contributions respectively from $t$- and $s$- channel
 $2 \leftrightarrow 2$ scattering processes ($\Delta L =1$)
involving the SM gauge fields \cite{hepph0310123}.
Similarly, $S^{(\text{quark})}_{H_t i}$
($W^{(\text{quark})}_{H_t i}$) and 
$S^{(\text{quark})}_{H_s i}$ 
($W^{(\text{quark})}_{H_s i}$) 
are contributions from $t$- and $s$- channel
$2 \leftrightarrow 2$ scattering processes ($\Delta L =1$)
involving the top quark.

For the total contributions of the $2 \leftrightarrow 2$ processes   
involving the SM gauge fields and the top quark
to the production of $N_i$ and to the wash-out terms
we get from Eqs. \eqref{eq:St} - \eqref{eq:Ws}:
\begin{eqnarray}
\label{eq:SG}
    S^{(\text{gauge})}_{i}  &=& 4\,S^{(\text{gauge})}_{A_t i} + 2\, S^{(\text{gauge})}_{A_s i}\,,\\
\label{eq:SQ}
    S^{(\text{quark})}_{i} &=& 4\,S^{(\text{quark})}_{H_t i} + 2\,D^{(\text{quark})}_{H_s i}\,,\\
\label{eq:WG}
    W^{(\text{gauge})}_i   &=& 4\,W^{(\text{gauge})}_{A_t i} + 2\,W^{(\text{gauge})}_{A_s i}\,,\\
\label{eq:WQ}
    W^{(\text{quark})}_i   &=& 4\,W^{(\text{quark})}_{H_t i}+ 2\,W^{(\text{quark})}_{H_s i}\,.
\end{eqnarray}
%

We take into account the thermal effects  
in the production of $N_{1,2}$ in Eqs. \eqref{BEsRLThN} and \eqref{BEsRLThL} 
using the results derived in \cite{1202.1288} for 
$D_i$, $S^{(\text{gauge})}_{i}$ and $S^{(\text{quark})}_{i}$
in the relevant case of relativistic $N_{1,2}$. 
As was shown in \cite{1202.1288}, the indicated three contributions
vary little in the interval of temperatures of interest, 
$T\sim (100 - 1000)$ GeV, and we have approximated them as 
constants equal to their respective average values in this interval. 
Adapting the results obtained in \cite{1202.1288} 
to the set-up utilised by us we get:
\begin{eqnarray}
\label{eq:DTh}
D_i                     &=& 0.232 \frac{\kappa_i}{z^2 K_2(z)}\,,\\
\label{eq:SQTh}
S^{(\text{quark})}_i    &=& 0.102 \frac{\kappa_i}{z^2 K_2(z)}\,,\\
\label{eq:SGTh}
S^{(\text{gauge})}_i    &=& 0.218 \frac{\kappa_i}{z^2 K_2(z)}\,.
\end{eqnarray}
%
where the parameter $\kappa_i$ is the ratio between the 
total decay rate of $N_i$ at zero temperature,
\begin{equation}  
\Gamma_{ii}\equiv (Y^\dagger Y)_{ii}\,M_i/ 8\pi\,,
\label{eq:Gammaii}
\end{equation}
%
and the Hubble expansion rate ($H$) at $z=1$. 
It proves convenient to cast $\kappa_i$ in the form:
\begin{equation}
\kappa_i\equiv\frac{\tilde{m}_i}{m_*}\,,
\end{equation}
%
with $m_*=(8\pi^2v^2/3M_p)\sqrt{(g_*\pi)/5}\approx10^{-3}$ eV, 
$M_P$ being the Planck mass.

Using the generic relations for the wash-out terms,
\begin{eqnarray}
    W_i^D       &=& \frac{2}{3} D_i N_{N_i}^{eq},\\
    W_i^{s,\,t} &=& \frac{2}{3} S^{s,\,t}_i N_{N_i}^{eq},
\label{eq:WDst}
\end{eqnarray}
%
we get:
\begin{eqnarray}
\label{eq:WDth}
    W^D_i                   &=& 0.058 \; \kappa_i,\\
 \label{eq:WQth} 
    W^{(\text{quarks})}_i   &=& 0.0255 \; \kappa_i,\\
 \label{eq:WGth}   
    W^{(\text{gauge})}_i    &=& 0.0545 \; \kappa_i.
\end{eqnarray}
%
The sum of the three terms is compatible with the result 
obtained in \cite{1403.2755}.

We stress that Eqs. \eqref{eq:DTh} - \eqref{eq:SGTh}
and \eqref{eq:WDth} - \eqref{eq:WGth} are valid only for $z < 1$. 
Moreover, as $z^2 K_2(z) \simeq 2$ for $z\ll 1$, all the terms given in Eqs. 
\eqref{eq:DTh} - \eqref{eq:SGTh} and \eqref{eq:WDth} - \eqref{eq:WGth}
are basically constant at $z\ll 1$. The behaviour 
at $z > 1$ is not relevant for our analysis 
as we are considering $M \leq 100$ GeV, so that the 
production of the baryon asymmetry stops at $z_{\rm sph} < 1$.

%
\subsection{The CP Violating Asymmetry}
%

 We define the CP violating (CPV) asymmetry with the inclusion of 
thermal effects as in \cite{1606.00017}, 
but taking also into account the flavour effects, namely 
\cite{Dev:2014laa,Dev:2014wsa,Bambhaniya:2016rbb} 
\footnote{The CP-asymmetry is defined with two flavour indices because 
in the quantum treatment
and in certain regimes (e.g., GUT scale leptogenesis) 
the off-diagonal terms are also relevant (see, e.g., \cite{Moffat:2018wke}).
}:
\begin{equation}
\label{CP}
\epsilon^{(i)}_{\alpha\alpha}=\sum_{i\not=j}
{\rm sgn}(M_i-M_j)\,
I_{ij,\alpha\alpha}\,\frac{2x^{(0)}\gamma(z)}
{4\frac{\Gamma_{22}}{\Gamma_{jj}}(x^{(0)}+x_T(z))^2+\frac{\Gamma_{jj}}{\Gamma_{22}}\gamma^2(z)}\,,
\end{equation}
%
where
\begin{equation}
\label{YukawaPiece}
I_{ij,\alpha\alpha}=\frac{{\rm Im}\left[Y^{\dagger}_{i\alpha}Y_{\alpha j}\left(Y^{\dagger}Y\right)_{ij}\right]
+\frac{M_i}{M_j}{\rm Im}\left[Y^{\dagger}_{i\alpha}Y_{\alpha j}\left(Y^{\dagger}Y\right)_{ji}\right]}{\left(Y^{\dagger}Y\right)_{ii}\left(Y^{\dagger}Y\right)_{jj}}\,.
\end{equation}
%
In Eq. \eqref{CP} the quantity 
$x^{(0)}\equiv \Delta M^{(0)} / \Gamma_{22}$, 
$\Delta M ^{(0)}$ being the $N_2 - N_1$ mass splitting at zero temperature 
\footnote{
The first term in the numerator of 
the expression in Eq. (\ref{YukawaPiece}), as can be shown, 
is lepton number violating (LNV), while the second term is lepton 
number conserving (LNC). Our numerical analyses show that 
under the conditions of the ``freeze-out'' leptogenesis mechanism 
we are studying, the dominant contribution in the generation of the baryon 
asymmetry compatible with the observations 
is given by the LNV term, with the LNC term 
giving typically a subdominant contribution; 
in certain specific cases the LNC contribution 
is of the order of, but never exceeds, the LNV one.
}.
The term that multiplies ${\rm sgn}(M_i-M_j)\,I_{ij,\alpha\alpha}$ 
in Eq. \eqref{CP} is due to heavy-neutrino mixing effects 
\footnote{If the CPV asymmetry $\epsilon^{(i)}_{\alpha\alpha}$
 is produced in $N_{1,2}$ decays one has to include in
 $\epsilon^{(i)}_{\alpha\alpha}$ also the effect of  
 heavy-neutrino oscillations \cite{Dev:2014wsa,Bambhaniya:2016rbb} 
 (which is different from the 
 the effect dominating the oscillation leptogenesis scenario 
  \cite{Akhmedov:1998qx,Asaka:2005pn}).
 Here we consider the case when the CPV asymmetry 
 $\epsilon^{(i)}_{\alpha\alpha}$ 
 is generated in the Higgs decays,  
 $\Phi^{(-)}\rightarrow l^{-} + N_j$, $ \Phi^{(+)}\rightarrow l^{+} + N_j$.
}.
Thermal corrections to the $N_2 - N_1$ mass splitting, $\Delta M_T$, 
with the total mass splitting given by   
$\Delta M = \Delta M^{(0)} + \Delta M_T$, 
are relevant in the denominator of the expression for 
$\epsilon^{(i)}_{\alpha\alpha}$ only and are 
accounted for by the term $x_T(z)$ \cite{1606.00017}: 
\begin{equation}
x_T(z)\equiv \frac{\Delta M_T(z)} {\Gamma_{22}}\simeq\frac{\pi}{4z^2}\sqrt{\left(1-\frac{\Gamma_{11}}{\Gamma_{22}}\right)^2+
4\frac{|\Gamma_{12}|^2}{\Gamma_{22}^2}}\,,
\end{equation}
%
where $\Gamma_{ij}\equiv (Y^\dagger Y)_{ij} \sqrt{M_iM_j} / 8\pi$.
The function $\gamma(z)$ in Eq. \eqref{CP} quantifies the thermal effects 
to the $N_{1,2}$ self-energy cut \cite{1606.00017} and is 
determined by
\begin{equation}
    \gamma(z) \equiv \left\langle \frac{p_\mu L^\mu}{p_\nu q^\nu} \right\rangle\,.
\end{equation}
%
Here $p$ and $q$ are the charged lepton and heavy Majorana neutrino 
four-momenta respectively, $L$ is defined in~\cite{1211.2140} 
and the angular brackets indicate a thermal average. 

 The function  $\gamma(z)$ depends on the masses of the 
Higgs boson $M_H(T)$, charged leptons $M_l(T)$ and 
heavy Majorana neutrinos $M_{1} \cong M_2 = M$ (it does not depend on the 
mass splitting $\Delta M$).
At $ T > T_{\rm EW} \cong 160$ GeV, $T_{\rm EW}$ being the temperature 
at which the electroweak phase transtion (EWPT) sets in 
\cite{DOnofrio:2014rug}, 
the Universe is in the symmetric phase and 
the Higgs vacuum expectation value is zero. 
The Higgs boson and the charged leptons possess only thermal
masses 
\footnote{The thermal corrections to the masses of the 
heavy Majorana neutrinos $N_{1,2}$ are negligibly small 
\cite{hepph0310123,Anisimov:2010gy}.
}.
For the charged lepton thermal masses we use the expression 
given in \cite{Anisimov:2010gy}:
\begin{equation}
M_l(T) \cong \dfrac{1}{4} \sqrt{3\,g^2 + (g')^2}\,T\,,
\label{eq:MlT}
\end{equation}
%
where $g = 0.65$ and $g' = 0.35$ are respectively the Standard Model 
$SU(2)_L$ and $U(1)_{\rm Y_W}$ gauge coupling constants.
At  $T < T_{\rm EW}$, the Higgs VEV
$v(T)$ grows approximately as 
(see, e.g., \cite{DOnofrio:2014rug,1606.00017}): 
$v^2(T) = (1 - T^2/T^2_{\rm EW})\,v^2\Theta(T_{\rm EW} - T)$, 
where $v \equiv v(T=0) \cong 246$ GeV is the VEV value at zero temperature.
Correspondingly, the charged lepton mass $M_l$ receives 
a non-zero contribution $M_l(v(T))$ 
in the interval $T_{\rm sph} \leq T < T_{\rm EW}$
due to $v(T) \neq 0$:
$M^2_l = M^2_l(v(T)) + M^2_l(T)$, $l=e,\mu,\tau$.
The EWPT contribution under discussion $M_l(v(T))$ 
is proportional to the zero temperature experimentally determined 
mass  $m_l$ of the charged lepton $l$: 
$M_l(v(T)) = m_l\,v(T)/v$.
It is not difficult to convince oneself that for 
$T$ in the interval  $T_{\rm sph} \leq T < T_{\rm EW}$ one has 
$M^2_l(v(T)) \ll M^2_l(T)$. Thus, for 
$T \geq T_{\rm sph}$ of interest the charged 
lepton masses are given by their thermal 
contributions specified in Eq. (\ref{eq:MlT}).

For the  Higgs mass $M_H(T)$ we employ the results 
obtained from the thermal effective potential 
given, e.g., in \cite{Quiros:1999jp,Senaha:2020mop},
which takes into account the effects of the EWPT 
in the interval of temperatures 
$T_{\rm sph} \lesssim T \lesssim T_{\rm EW}$, 
$T_{\rm sph} = 131.7$ GeV being the sphaleron decoupling 
temperature \cite{DOnofrio:2014rug}.
{The discussion of the behaviour of $M_H(T)$ in the interval 
of temperatures of interest is outside the scope of the present study;
it can be inferred from the aforementioned EWPT effective potential.
We give here only the expression for the thermal 
contribution to the Higgs mass:
\begin{equation}
M^{\rm therm}_H(T) \cong \dfrac{1}{4} 
\sqrt{3\,g^2 + (g')^2 + 4\,h^2_t + 8\,\lambda}\,T\,.
\label{eq:MHT}
\end{equation}
%
Here $h_t$ and $\lambda$ are the top Yukawa coupling and the 
Higgs quartic coupling, respectively. 
In the numerical analysis which follows we use 
$h_t = 0.993$ and $\lambda = 0.129$, which correspond to the 
top and zero temperature Higgs masses of 172.76 GeV and 125 GeV
and Higgs VEV of 246 GeV.

 With the chosen values of the couplings we have 
\begin{equation}
 M_l(T) \cong 0.296\,T\,,~~~~M_H(T) \cong 0.632\,T\,.
\label{eq:MlTMHT}
\end{equation}
%
It is easy to check using $M_l(T)$ and $M_H(T)$
given  in the preceding equation that for the values of 
$z$ lying approximately in the interval
$0.34 \lesssim z \lesssim 0.93$
the Higgs and heavy Majorana decay processes 
are kinematically forbidden  \cite{1606.00017,Anisimov:2010gy}.
For $z\lesssim 0.34$ the only allowed processes are Higgs decays 
to heavy Majorana neutrinos and charged leptons, whereas at larger 
$z \gtap 0.93$ only the heavy Majorana neutrino decays are allowed. 
This is reflected in the behavior of the function 
$\gamma(z)$ shown in Fig. \ref{fig:gamma}: 
for $0.34 \lesssim z \lesssim 0.93$ one has $\gamma(z) = 0$. 
We find also that for 
$z \ll 1$, $\gamma(z) \cong 23.5$ and for $z \gg 1$, $\gamma(z) \cong 1$.

 For $T > T_{\rm sph}$ of interest, 
the charged lepton masses satisfy 
$M_L(T) >  0.296T_{\rm sph} \cong 39$ GeV.
This implies that in order for the decays of the heavy Majorana neutrinos 
$N_{1,2}$ to be in principle kinematically possible at  $T > T_{\rm sph}$, 
the masses of $N_{1,2}$ must satisfy 
$M_{1,2} > M_L(T_{\rm sph}) \cong 39$ GeV. 
Taking into account also the Higgs mass leads obviously to a larger
lower bound on $M_{1,2}$ (see, e.g.,  \cite{1606.00017}).

 We note further that in the interval of temperatures 
$T_{\rm sph} \leq T < T_{\rm EW}$, the thermal contribution 
to the masses of the SM top quark, Higgs, $W^\pm$ and $Z$ bosons 
are all of the order of $\tilde{g}T$, $\tilde{g}$ being 
one of the SM couplings $g$, $g'$, $h_t$ and 
$\lambda$, and that the contribution to these masses 
of the non-zero temperature dependent Higgs vacuum 
expectation value $v(T)$, determined earlier, is of the same 
order. At the same time, the momenta of the particles in the thermal 
bath are of the order of $\pi T$ and are much larger than the masses, 
so all the particles relevant for our discussion 
are ultrarelativistic at the temperatures on interest  
\cite{Ghiglieri:2016xye}. This implies also, in particular, 
that the expressions for the decay and scattering terms 
introduced in the preceding subsection are valid actually for 
$T \geq T_{sph}\cong 131.7$ GeV.

  When collinear emissions of soft gauge bosons, 
present in the thermal bath of the Universe at the epoch 
of interest, are also included in the decay processes, 
the disallowed region 
discussed earlier 
becomes accessible to the Higgs decays 
due to the increased range of kinematic possibilities 
\cite{Anisimov:2010gy}.
In the calculations performed by us we estimate the effects of 
these emissions by adding an interpolation of  $\gamma(z)$
across the ``gap'' interval $0.34 \lesssim z \lesssim 0.93$ 
\footnote{The effects of collinear emissions of soft gauge bosons
are included also in the decay terms $D_i$ 
and $W_i^D$ given in Eqs. (\ref{eq:DTh}), (\ref{eq:WDst})
and (\ref{eq:WDth}), as discussed in the preceding section.
}.
This is illustrated in Fig. \ref{fig:gamma} showing 
$\gamma(z)$ as a function of $z$. For a given value 
of $M$, the behaviour of $\gamma(z)$ at 
$z \geq z_{sph} \equiv M/T_{\rm sph}$ (or at $T \leq T_{\rm sph}$) 
is not relevant and should be ignored since the sphalerons 
decouple at $T_{\rm sph}$.

 We note that at high temperatures $\gamma(z)$ is sensitive to 
precise values of Higgs and charged lepton thermal masses. As we have 
indicated, in our numerical analysis 
we use Higgs and charged lepton thermal masses in agreement with 
that of~\cite{Anisimov:2010gy}. 
We estimate that a different choice of 
thermal masses, as well as a different interpolation of $\gamma(z)$, 
would only affect our final results by 
an insignificant factor.
\begin{figure}
\centering
\includegraphics[width=\textwidth]{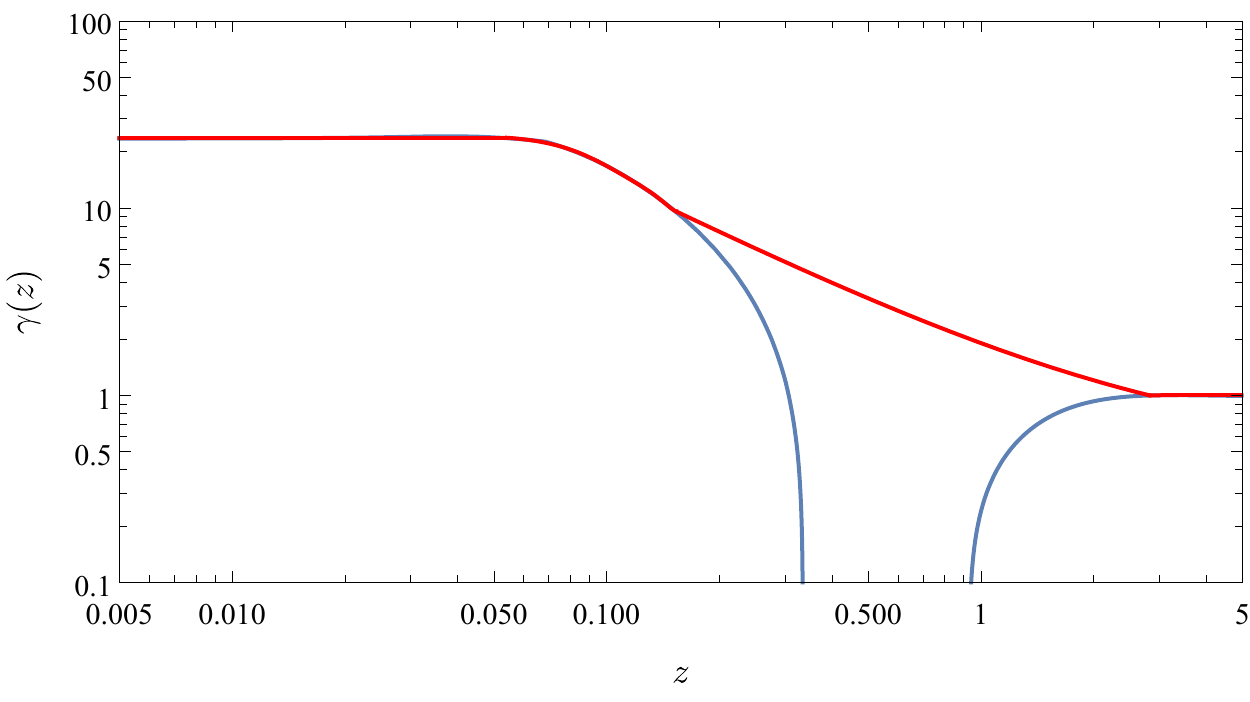}
\caption{ The function $\gamma(z)$. 
The blue line is obtained by taking into account the thermal 
masses of the Higgs boson and charged leptons 
as in Eq. \eqref{eq:MlTMHT}, 
while the red line shows the effect of accounting for 
the collinear emissions of soft gauge bosons  
(present in the thermal bath) 
in the Higgs decays  of interest. 
See text for further details.
}
\label{fig:gamma}
\end{figure}

 In the absence of thermal effects, namely setting $\gamma(z)=1$ and
$x_T(z)=0$, the second term in Eq. \eqref{CP} is maximised 
for $x^{(0)} \simeq 0.5$.
This is the ``resonant" behaviour 
typical to RL without thermal corrections. When thermal corrections are 
taken into account, it is not possible to choose one value of $x^{(0)}$ 
for which we have resonance at all temperatures. For this reason, 
the results we will show further are given for different 
values of $x^{(0)}$.

For $e^{|\xi|} \gg e^{-|\xi|}$, the term involving the Yukawa couplings 
given in Eq. \eqref{YukawaPiece} 
is proportional to $\sin(2\omega)e^{-2|\xi|}$. 
Hence, large values of $|\xi|$ suppress the CPV asymmetry, 
while $\omega = (2n+1)\pi/4$ maximises 
it (in absolute value). For $|\xi|\lesssim 1$, a slightly different dependence 
on $\omega$ appears in both the Yukawa coupling term (in the denominator) and 
in the mixing term of Eq. \eqref{CP}, so that the maximal value of the 
CPV asymmetry is actually reached for different values of 
$\omega$ (depending on $\xi$). 
We find, however, that a more precise choice of $\omega$ would not lead 
to significant differences in the BAU. Therefore, 
in obtaining
our results we set $\omega = \pi/4$ or $3\pi/4$ (to match the sign of the BAU) 
in order to maximise the CPV asymmetry at large values of $|\xi|$ 
\footnote{For such choice of $\omega$ and provided $\Delta M / M \ll 1$, 
we find that $\Gamma_{11}/\Gamma_{22}\sim 1$ for any $\xi$. Given this and 
summing the CPV asymmetry in Eq. \eqref{CP} over flavours and RH neutrino 
indices, we get the same form of the asymmetry used in 
\cite{1606.00017}, with a factor of 2 that accounts for the two RH 
neutrinos present in our analysis. 
}. 
In the analysis which follows we consider only values of $\xi \geq 0$ 
since the results are symmetric for the corresponding $\xi < 0$.

Throughout we use the ULYSSES \cite{Granelli:2020pim} Python package for 
numerical solutions of the Boltzmann equations. 
We will present results of the numerical analysis for 
the Majorana phase  $\alpha_{23}$ set to zero, 
adding comments in certain cases on how the results change 
for $\alpha_{23}$ having a non zero value in the interval 
$0 < \alpha_{23} \leq 4\pi$. 

%
\subsection{Thermal Initial Abundance}
\label{sec:TIA}
%

Consider the case of thermal initial abundance (TIA) of the 
heavy Majorana neutrinos, 
$N_{N_i}(z_0) = N_{N_i}^{eq}(z_0)$. We can set the ratio 
$N_{N_i}/ N_{N_i}^{eq} = 1$ 
in the r.h.s. of Eq. \eqref{BEsRLThL} since under 
the indicated initial condition the deviations of $N_{N_i}$ from  
$N_{N_i}^{eq}$ for any $z > z_0$ 
of interest for our analysis
are sufficiently small and can be neglected.
For the sum of three wash-out factors, $W_i$, in this case we get:
\begin{eqnarray}
\label{eq:TIAW}
    W^\text{TIA}_i &\equiv& W^D_i +  W^t_i +  W^s_i\,\\
        &=& W^D_i + W^{(\text{quarks})}_i 
+  W^{(\text{gauge})}_i \cong 0.138\;\kappa_i\,.
\end{eqnarray}
%
The flavoured washout 
terms in Eq. (\ref{BEsRLThL}) are given by  
$W_{i \alpha} \equiv p_{i \alpha} W^\text{TIA}_i$. 
\begin{figure}[t!]
\centering
\includegraphics[width=\linewidth]{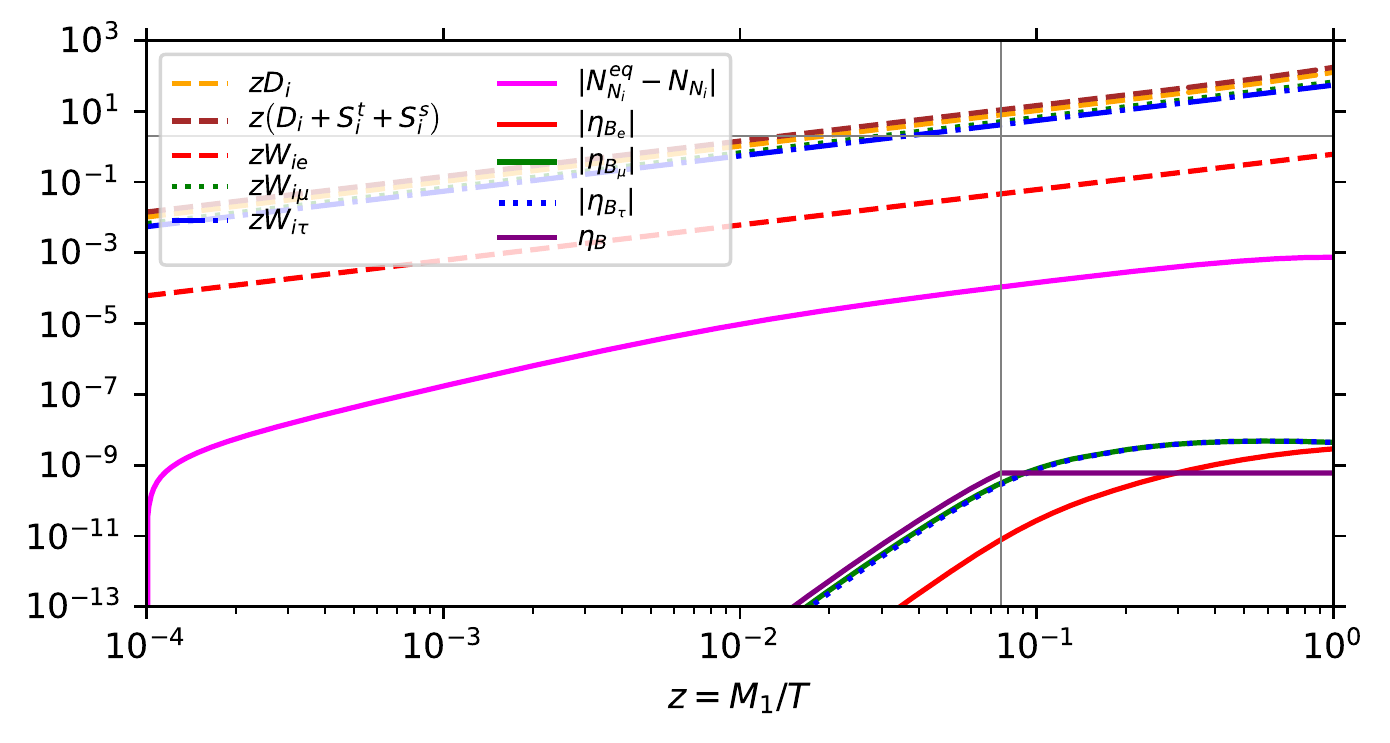}
\caption{
The evolution of the lepton flavour and baryon asymmetries,
of $|N_{N_i}^{eq}-N_{N_i}|$
and of the corresponding decay, 
scattering and washout rates that govern the evolution of the asymmetries 
in Eqs. \eqref{BEsRLThN} and \eqref{BEsRLThL} in the case of TIA of $N_{1,2}$. 
The figure is obtained for $\delta = 3\pi/2$, $M = 10$ GeV, 
$x^{(0)} = 100$ and $\xi = 2.05$.
The vertical grey line is at $z_{sph} = 0.076$ and is the endpoint of 
evolution of the baryon asymmetry 
$\eta_B$. 
The horizontal grey line 
at $2$ is roughly indicating where the different processes 
get into equilibrium. See text for further details.
}
\label{fig:TIA10GeV}
\end{figure}
%
Due to the projection probability $p_{i\alpha}$ 
the wash-out terms exhibit strong flavour dependence.

 In what follows we discuss the results of our numerical analysis.
Figure \ref{fig:TIA10GeV} shows the evolution of the lepton and baryon 
asymmetries, the difference $|N_{N_i}^{eq}-N_{N_i}|$, the decay, scattering and 
wash-out rates for $\delta = 3\pi/2$, $M=10$ GeV, $x^{(0)} = 100$ and 
$\xi = 2.05$ -- the maximal value of $\xi$ for which we can 
have successful leptogenesis 
for  $M=10$ GeV -- and $x^{(0)} = 100$.
The baryon asymmetry $\eta_{B l}$ originates from 
the CPV asymmetry in the lepton charge (flavour)
$L_l$ ($l$), $l=e,\mu,\tau$.
Thus, the total baryon asymmetry is 
$\eta_B = \eta_{Be} + \eta_{B\mu} + \eta_{B\tau}$. 
The figure illustrates the typical scenario of ``freeze-out" 
leptogenesis, namely, the case when the departure from equilibrium of 
$N_{N_i}(z)$ is what drives the generation of the lepton (and baryon) 
asymmetry. The total baryon asymmetry, to which all flavour CPV 
asymmetries contribute, freezes at $z_{sph} =  M/T_{sph} \cong 0.076$, 
$T_{sph} = 131.7$ GeV being the sphaleron decoupling temperature, 
which is marked by the vertical grey line in the figure.

 For the choice of parameters in the figure, 
the asymmetries  $|\eta_{B\mu}|$ and $|\eta_{B\tau}|$ 
exhibit almost identical evolution 
for $z \leq z_{sph}$  
and are by a factor $\sim 100$ larger than 
$|\eta_{B e}|$. This difference reflects
the difference between 
$\epsilon_{\mu\mu}^{(i)}$ ($\epsilon_{\tau\tau}^{(i)}$) and $\epsilon_{ee}^{(i)}$. 
Thus, in this case, 
$\eta_B \cong \eta_{B\mu} + \eta_{B\tau}$.
The fact that 
$|\eta_{B\mu}| \cong |\eta_{B\tau}| \gg |\eta_{Be}|$ 
can have important implications in what concerns the 
possibility of wash-out of the baryon asymmetry 
by lepton number non-conserving effective operators 
of dimension higher than four 
that might be ``active'' at the energy scales of interest 
\cite{Deppisch:2013jxa,Deppisch:2017ecm}.  
For $z > z_{sph}$, and therefore after sphaleron freeze-out, 
the asymmetry $|\eta_{B e}|$ converges to the asymmetries 
$|\eta_{B\mu}|$ and $|\eta_{B\tau}|$.
Qualitatively similar behaviour is seen for a range of 
$x^{(0)}$ and $M$ values, with the main difference being the overall 
scale of the asymmetry evolution.

To understand the impact of flavour effects, we compare 
the obtained results with the results in the unflavoured case. The unflavoured 
approximation is equivalent to taking in Eq. \eqref{BEsRLThL} 
$p_{i\alpha} = 1$ for every $\alpha$ and then sum over all the flavours. 
It roughly corresponds to taking the total asymmetry to be the asymmetry 
in the dominant flavour (either muon or tauon in 
in the considered
case). As it is then clear from Fig. \ref{fig:TIA10GeV}, 
this approximation would 
only lead to a $\mathcal{O}(2)$ difference in the value of $\eta_B$.
A more detailed analysis shows that 
flavour effects in the TIA case lead, in general, to a moderate 
enhancement by a factor of $\sim (2-3)$ of the baryon asymmetry.

 We show in Fig. \ref{fig:xiMassTIA} the maximal values of $\xi$, 
for which we can have successful RL as a function of the mass scale $M$. 
We recall that to the maximal values of $\xi$ 
there correspond maximal values of $\sum|(RV)|^2$ (see eq. \eqref{RV}).
The curves of different colours correspond to the contours at 
different $x^{(0)}$ for which the maximal baryon asymmetry coincides 
with the present observed value of $\approx 6\times 10^{-10}$.
In the blue region, the 
baryon asymmetry is too small compared to 
the observed value. In the white region instead, 
we can always vary $x^{(0)}$, $\omega$ and/or the 
CPV phases in the PMNS matrix
so to get the correct value for $\eta_B$.

In the considered scenario, for a given $x^{(0)}$, the lower the mass, 
the less is the time for the system 
to depart from equilibrium before $z_{sph}$, and so greater must be 
the CPV asymmetry, and slower the processes keeping $N_i$ in equilibrium, 
so that the baryon asymmetry freezes at the observed value.  Correspondingly, 
by lowering the mass, the maximal value of $\xi$ for which we can 
have successful RL decreases. 
This leads to a lower bound on the mass $M$ that depends on $x^{(0)}$, 
for which RL can be successful.
In the region of small $\xi$ the dependence on $\xi$ of the CPV asymmetry 
and wash-out terms is less trivial and strongly dependent on the 
leptonic CPV phases, leading to the growth with the mass, 
as shown in Fig. \ref{fig:xiMassTIA}.

As long as $x^{(0)}\lesssim x_T(z)$, the CPV asymmetry grows 
with $z$ and $x^{(0)}$, whilst for $x^{(0)}\gtrsim x_T(z)$ the asymmetry 
is constant ($\gamma(z)$ is constant at $z\ll 1$) and is suppressed by 
large values of $x^{(0)}$. As a consequence of this behaviour, 
we find that the minimal lower bound on the mass $M$ 
for having successful flavoured RL 
is reached for $x^{(0)}\approx 10^3$ and reads $M\cong 5$ GeV.
\begin{figure}[t]
\centering
\includegraphics[width=\textwidth]{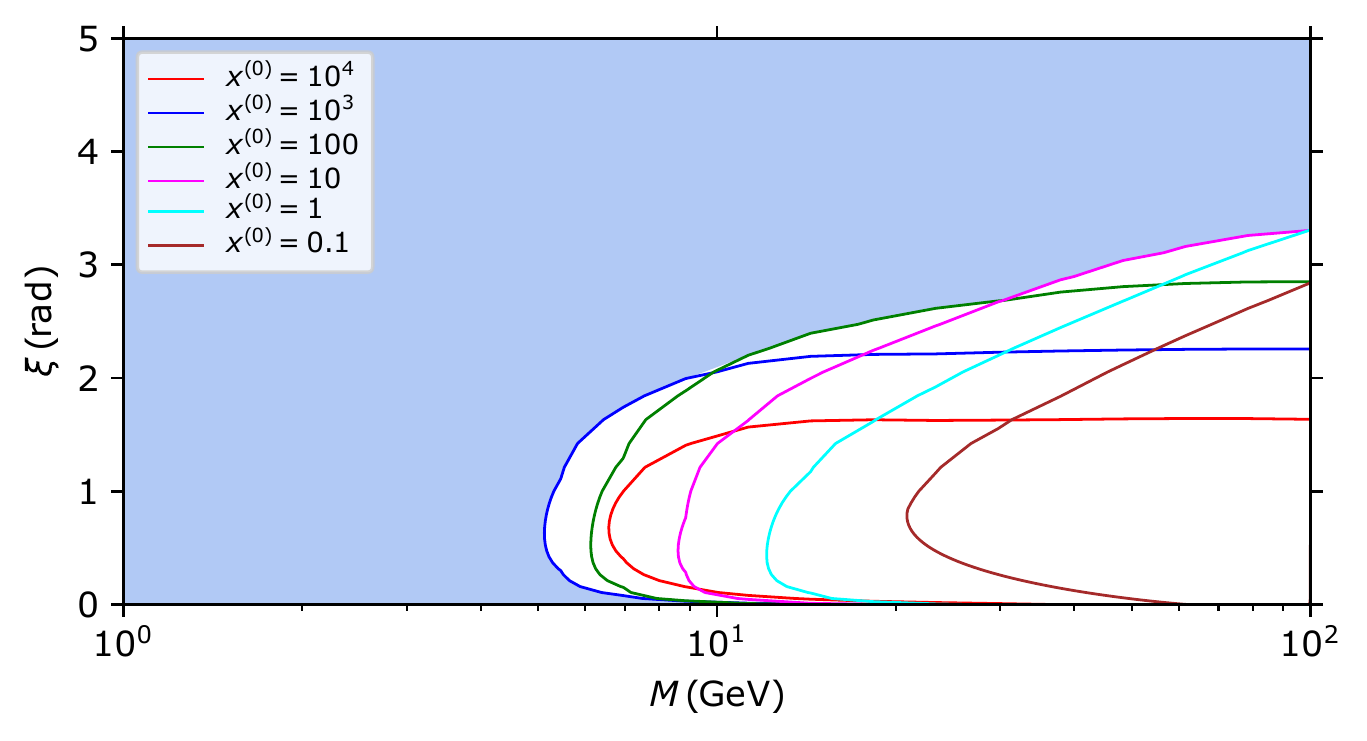}
\caption{
The allowed region (white) in the $\xi-M$ plane, 
for which we can always have successful RL by varying $x^{(0)}$, $\omega$ 
and/or the CPV phases in the PMNS matrix,
in the case of TIA. 
For values in the blue region the baryon asymmetry is always too small 
compared to that observed today. 
The solid contours corresponds to the maximal value of $\xi$
and $x^{(0)} = 10^4$ (red), $10^3$ (blue), $10^2$ (green), 
10 (magenta), 1 (cyan) and 0.1 (brown),
for which the predicted asymmetry coincides with the observed one. 
The figure is obtained for $\delta = 3\pi/2$.
See text for further details.
}
\label{fig:xiMassTIA}
\end{figure}
%

Given the mass $M$ of the heavy Majorana neutrinos $N_{1,2}$ 
($M_{1,2}\cong M$, $M_2 - M_1 \equiv \Delta M \ll M$), 
the values of the parameter $\xi$ determine, as we have already 
indicated, the magnitude of the charged and neutral current 
couplings of $N_{1,2}$, $(RV)_{lj}$, in the weak interaction Lagrangian. 
To the maximal values of $\xi$ there correspond maximal values of 
$\sum_{l,i}|(RV)|^2$ (see eq. \eqref{RV}).
For the maximal value of  $\xi$ in the TIA case 
we find $\xi\cong 3.3$. It is reached for $M\cong 100$ GeV and 
$x^{(0)} \cong 1-10$. In the case of $M\cong 20$ GeV we get 
${\rm max}(\xi) \cong 2.5$.
Thus, for $M\cong 20$~(100) GeV we find 
${\rm max} (\sum_{l, i}|(RV)_{li}|^2) 
\cong 2.2\times 10^{-10}~(2.2\times 10^{-10})$.

The results presented in Fig. \ref{fig:xiMassTIA} 
have been obtained for the Majorana phase $\alpha_{23} = 0$.
We have investigated numerically how these results change 
when  $\alpha_{23} \neq 0$. For this purpose we have obtained versions of 
Fig.  \ref{fig:xiMassTIA} for 
$\delta = 270^\circ$ and $\alpha_{23} = 90^\circ,\,180^\circ,\,270^\circ,\, 
360^\circ,\,450^\circ,\,540^\circ$ and $630^\circ$.
Comparing the results for $\alpha_{23} = 0$ shown in  Fig. \ref{fig:xiMassTIA} 
with those derived for $\alpha_{23} \neq 0$ we do not find any significant 
dependence on the value of the Majorana phase $\alpha_{23}$.
We can conclude, in particular, that:\\ 
i) the minimal lower bound on the mass $M$ 
for having successful flavoured RL 
is practically the same, namely, $M\simeq 5$ GeV, 
for all values of $\alpha_{23}$ considered;\\
ii) the maximal value of $\xi$ is the same as found for $\alpha_{23} = 0$, 
namely, $\xi\cong 3.3$, and similarly to the $\alpha_{23} = 0$ case, 
is reached for $M\cong 100$ GeV;\\
iii) for the maximal value of $\xi$ reached at $M = 20$ GeV  
we find $\xi = 2.63$ at $\alpha_{23} = 360^\circ$; it is somewhat 
larger than the value of $\xi = 2.55$ found for $\alpha_{23} = 0$ 
and corresponds to ${\rm max} (\sum_{l, i}|(RV)_{li}|^2) 
\simeq 2.8\times 10^{-10}$.

Our detailed analysis showed that the wash-out factor for the asymmetry 
in the $e$-lepton charge and the related baryon asymmetry 
$\eta_{Be}$ exhibit weak dependence of $\alpha_{23}$,
while the wash-out factors for the asymmetries 
in the  $\mu$- and $\tau$- lepton charges and, 
correspondingly, the flavour baryon asymmetries 
$\eta_{B\mu}$ and $\eta_{B\tau}$, change significantly 
with $\alpha_{23}$. However these changes are 
``anti-correlated'' in the sense that the sum 
$\eta_{B\mu} + \eta_{B\tau}$ remains practically 
constant when $\alpha_{23}$ is varied.
As a consequence, also the total baryon asymmetry 
$\eta_B = \eta_{Be} + \eta_{B\mu} + \eta_{B\tau}$ 
does not show any noticeable dependence of $\alpha_{23}$.

We compare next briefly the results obtained by us with those derived in 
\cite{Klaric:2020lov} in the TIA case. The results in the decay scenario 
in the TIA case obtained in \cite{Klaric:2020lov} 
are indicated graphically in Fig. 2 in \cite{Klaric:2020lov}.
The minimal value of $M\simeq 5$ GeV for successful leptogenesis 
we find in our analysis is similar 
to the value of $\sim 7$ GeV found in \cite{Klaric:2020lov}.
In what concerns  $\sum_{l, i}|(RV)_{li}|^2$ 
(which is equivalent to the parameter $|U|^2$ in Fig. 2 in 
\cite{Klaric:2020lov}), for the maximal value of this 
important for the phenomenology parameter 
at $M = 10$, $20$ and $100$ GeV we find, 
respectively,  
${\rm max} (\sum_{l, i}|(RV)_{li}|^2)\simeq 2.1\times 10^{-10}$, 
$2.8\times 10^{-10}$ and $2.2\times 10^{-10}$, while, 
according to their Fig. 2, the authors of \cite{Klaric:2020lov} get 
 $\sim 5\times 10^{-10}$, $\sim 1.0\times 10^{-9}$ and $\sim 1.2\times 10^{-9}$.
Thus, our results for the maximal value of the observable 
$(\sum_{l, i}|(RV)_{li}|^2)$
at $M=10$, $20$ and $100$ GeV are somewhat smaller that those 
obtained in \cite{Klaric:2020lov} and 
thus can be considered as more conservative.  

 For completeness we summarise our results for the observables 
 ${\rm max} (\sum_{l, i}|(RV)_{li}|^2)$  and 
${\rm min} (\sum_{l, i}|(RV)_{li}|^2)$ for a large set of values of  
$M = 10$, 20, 30, 40, 50, 60, 70, 80 and 85 GeV:
${\rm max} (\sum_{l, i}|(RV)_{li}|^2)\times 10^{10} 
\cong  2.08$, 2.83, 2.50, 2.68, 2.78, 2.83, 2.74, 2.59 and 2.49, 
while  
 ${\rm min} (\sum_{l, i}|(RV)_{li}|^2)\times 10^{12} 
\cong 5.89$, 2.94, 1.96, 1.47, 1.18, 0.98, 0.84, 0.735 and 0.692.
They correspond to $\alpha_{21} = 2\pi$ and $\delta = 3\pi/2$.
The quoted values of ${\rm min} (\sum_{l, i}|(RV)_{li}|^2)$ 
are in perfect agreement with the values reported in 
\cite{Klaric:2020lov}.


\subsection{Vanishing Initial Abundance}
\label{sec:VIA}
\begin{figure}[t!]
\centering
\includegraphics[width=\textwidth]{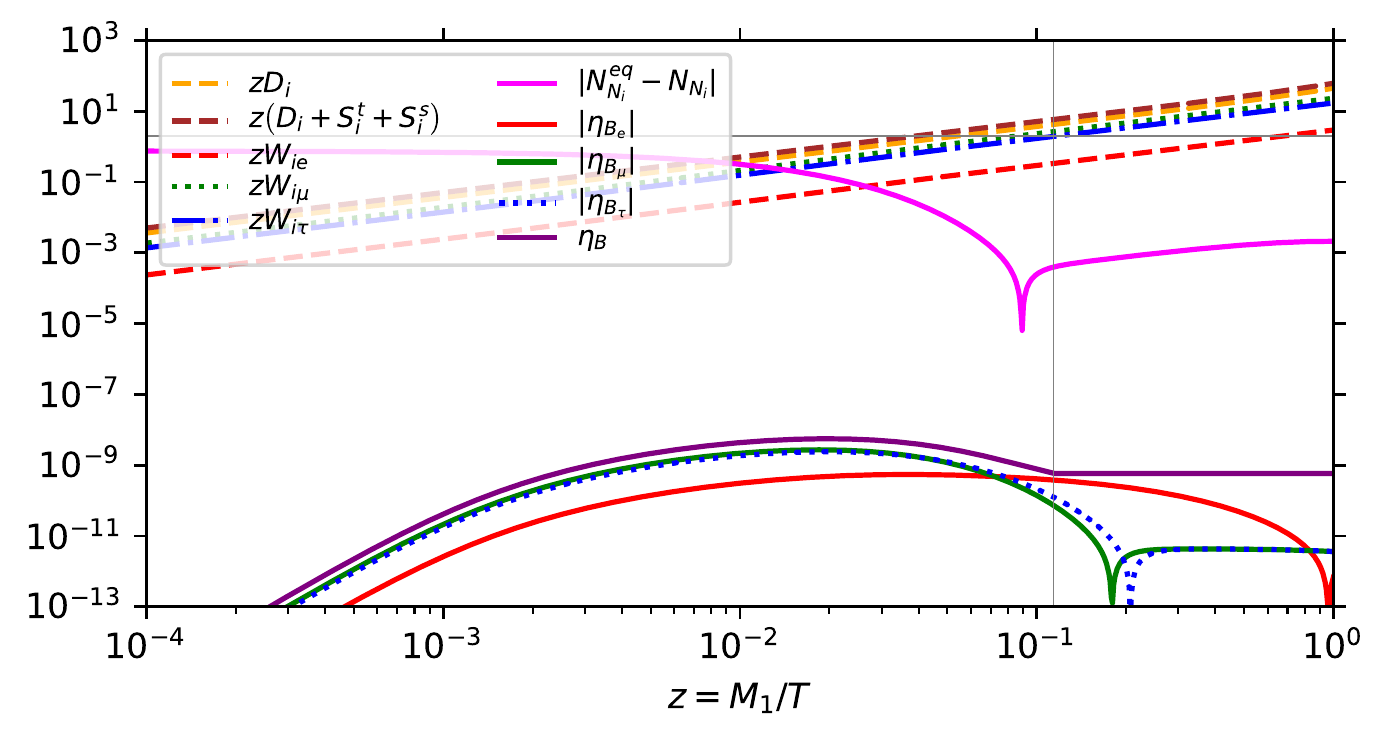}
\caption{
The evolution of the lepton-flavour and baryon asymmetries,
of $|N_{N_i}^{eq}-N_{N_i}|$
and of the corresponding decay, 
scattering and washout rates that govern the evolution of the asymmetries 
in Eqs. \eqref{BEsRLThN} and \eqref{BEsRLThL} in the case of vanishing 
initial abundance (VIA) of $N_{1,2}$. The figure is obtained for $\delta = 0$, 
 $M = 15$ GeV, $x^{(0)} = 10^6$ and $\xi = 1.53$.
The vertical grey line at $z_{sph} = 0.114$ is the endpoint 
of the evolution of the baryon asymmetry
$\eta_B$. The horizontal grey line at $2$ is roughly 
indicating where the 
different processes get into equilibrium. 
At $z\ll z_{sph}$, the asymmetry  $|\eta_{Be}|$ 
(solid red curve) is smaller in magnitude than the 
asymmetries $|\eta_{B\mu}|$ and $|\eta_{B\tau}|$.
However, the weaker washout of  $\eta_{Be}$
results in it dominating $\eta_{B\mu}$ and $\eta_{B\tau}$
by the time of sphaleron decoupling 
and in $\eta_B \cong \eta_{Be}$.
See text for further details.
}
\label{fig:VIA15GeV}
\end{figure}

\begin{figure}
\centering
\includegraphics[width=\textwidth]{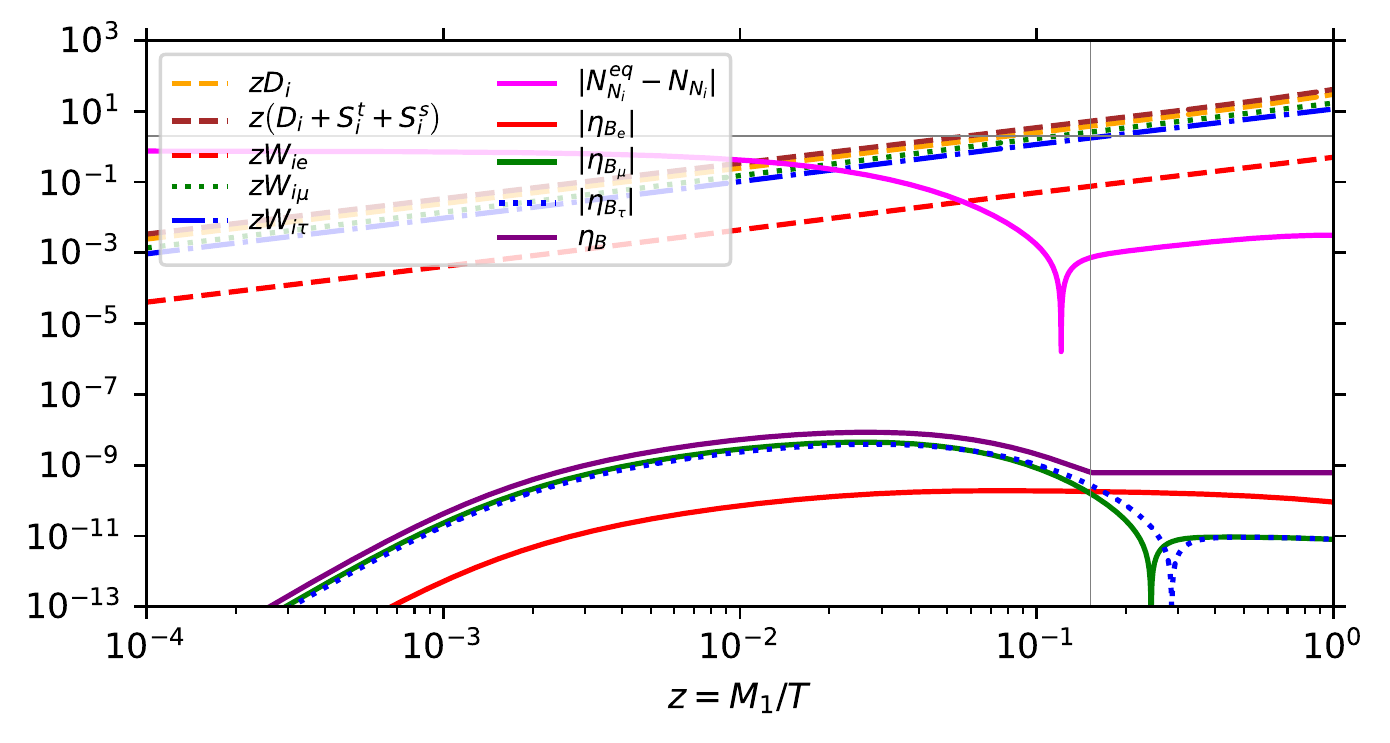}
\caption{ 
The same as in Fig. \ref{fig:VIA15GeV} but for  
 $\delta = 300^\circ$, $M = 20$ GeV, $x^{(0)} = 10^6$ and $\xi = 1.33$.
The vertical grey line at $z_{sph} = 0.15$ is the endpoint of the evolution 
of the baryon asymmetry $\eta_B$.
The horizontal grey line at $2$ is roughly indicating where the 
different processes get into equilibrium. 
In this case $\eta_B = \eta_{Be} + \eta_{B\mu} + \eta_{B\tau}$.
See text for further details.
}
\label{fig:VIA20GeV}
\end{figure}

\begin{figure}
\centering
\includegraphics[width=\textwidth]{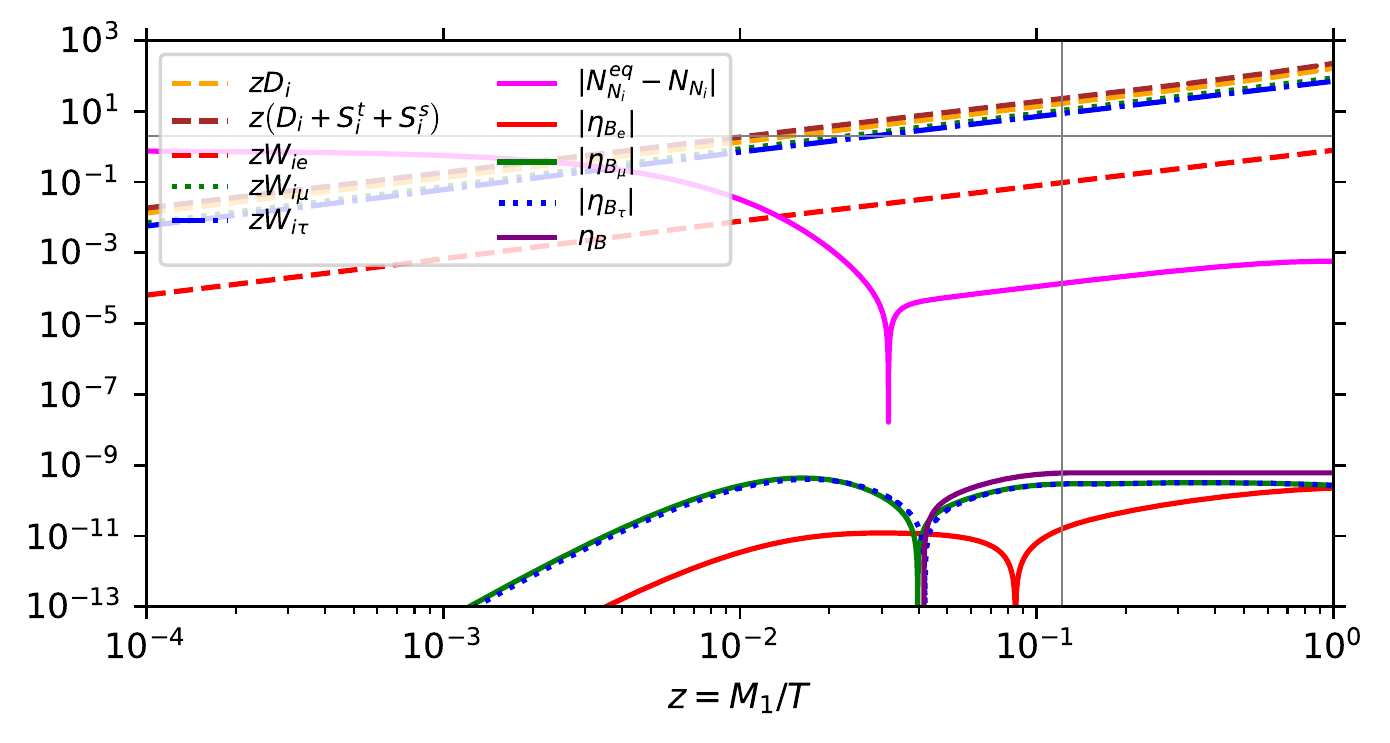}
\caption{
The same as in Fig. \ref{fig:VIA15GeV} but
for $\delta = 3\pi/2$, $M = 16$ GeV, $x^{(0)} = 10^3$ and 
$\xi = 2.18$.
The vertical grey line at $z_{sph} = 0.12$ 
is the endpoint of the evolution 
of the baryon asymmetry $\eta_B$.
The horizontal grey line at $2$ is roughly indicating where the 
different processes get into equilibrium. 
The figure illustrates a case
of  $\eta_B \cong \eta_{B\mu} + \eta_{B\tau}$.
See text for further details.
}
\label{fig:VIA16GeV}
\end{figure}

 We analyse next the case of a vanishing initial abundance (VIA) 
of $N_{1,2}$, i.e. $N_{1,2} \left(z_0\right) = 0$. For the wash-out terms, 
we assume on the basis of the results reported 
in \cite{hepph0310123} that 
\begin{eqnarray}
\label{eq:WGts}
    W^{(\text{gauge})}_{A_t i} \cong W^{(\text{gauge})}_{A_s i}\,,\\
\label{eq:WQts}
    W^{(\text{quark})}_{H_t i} \cong W^{(\text{quark})}_{H_s i}\,.
\end{eqnarray}
%
Under these conditions
\footnote{We have checked that choosing different relations 
between $W^{(\text{gauge})}_{A_t i}$ and $W^{(\text{gauge})}_{A_s i}$, 
and between $W^{(\text{quark})}_{H_t i}$ and $W^{(\text{quark})}_{H_s i}$,
does not lead to significant change of the results
obtained using Eqs. (\ref{eq:WGts}) and (\ref{eq:WQts}).
}
the wash-out term in Eq. \eqref{BEsRLThL} 
in the case of interest has the form:
\begin{equation}
\label{eq:WVIA}
\begin{split}
W^\text{VIA}_i &\equiv W^{D}_i + W^{t}_i  + \frac{N_{N_i}}{N_{N_i}^{eq}} W^{s}_i\\
&\cong \left (0.1113 + 0.0267\;\frac{N_{N_i}}{N_{N_i}^{eq}} \right)\, \kappa_i\,,
\end{split}
\end{equation}
%
The flavoured washout 
terms in Eq. (\ref{BEsRLThL}) are given by  
$W_{i \alpha} \equiv p_{i \alpha} W^\text{VIA}_i$. 

 In Figs. \ref{fig:VIA15GeV}, \ref{fig:VIA20GeV} 
and \ref{fig:VIA16GeV} we show 
the evolution of the leptonic asymmetries 
$N_{\Delta \alpha}$ respectively for 
i) $\delta = 0$,  $M = 15$ GeV ($z_{sph} = 0.114$),  
$x^{(0)} = 10^6$ and maximal $\xi = 1.53$,
ii) $\delta = 300^\circ$, $M = 20$ GeV ($z_{sph} = 0.15$), 
$x^{(0)} = 10^6$ and $\xi = 1.33$,
and iii) $\delta = 3\pi/2$,  
$M = 16$ GeV ($z_{sph} = 0.121$), $x^{(0)} = 10^3$ and
maximal $\xi = 2.18$, respectively. Also shown is the growth of 
$N_N$ towards the evolving equilibrium distribution $N^{eq}_N\left(z\right)$, 
governed by the combination $D_i + S^t_i + S^s_i$. The sphaleron transition 
occurs at $z_{sph}$ marked by the vertical grey line after which 
the baryon asymmetry $\eta_B$ is ``frozen" and remains constant 
at the value at $z_{sph}$. Sharp dips in the asymmetries correspond 
to sign changes as we always plot absolute values.

As is seen in Fig. \ref{fig:VIA15GeV}, the 
asymmetries $\eta_{B\mu}$ and $\eta_{B\tau}$, 
which are generated by the CPV $\mu$- and $\tau$- flavour asymmetries,
are strongly suppressed in the interval $0.07 \lesssim z \leq z_{sph}$
due to the relatively large wash-out factors. 
This is reflected in the sudden dips of the corresponding
curves as they are driven through zero by the 
wash-out  effects.
As a consequence, by the time of
sphaleron decoupling most of the 
baryon asymmetry is due to the lepton CPV asymmetry 
residing in the electron flavour, $\eta_B \cong \eta_{Be}$. 
Since $\eta_{Be}$ was mostly generated during the production of RH neutrinos, 
i.e., before $N_{Ni}$ reached $N^{eq}_{Ni}$, 
this case corresponds to the ``freeze-in" scenario 
of generation of baryon asymmetry.

We highlight the fact that, 
in contrast to the TIA case, in the VIA scenario 
illustrated in Fig. \ref{fig:VIA15GeV} flavour effects are crucial. 
Here, the ``dominant flavours'' are the muon and tauon, 
in the sense that both respective CPV asymmetries 
$\epsilon^{(i)}_{\mu \mu}$ and $\epsilon^{(i)}_{\tau \tau}$
and projection probabilities $p_{i \mu}$ and $p_{i \tau}$
are greater than the electron flavour ones,
$\epsilon^{(i)}_{e e}$ and  $p_{ie}$:
$|\epsilon^{(i)}_{\mu \mu}|,|\epsilon^{(i)}_{\tau \tau}| > 
|\epsilon^{(i)}_{e e}|$, $p_{i \mu},p_{i \tau} > p_{ie}$.
The unflavoured approximation would then neglect 
the electron CPV asymmetry and consequently $\eta_{Be}$,
which actually contributes most in the flavoured 
scenario. In this particular case, flavour effects lead to a 
$\mathcal{O}(300)$ enhancement with respect to the unflavoured case
\footnote{The enhancement can also come with a peculiar difference in sign, 
which reflects the fact that in this intermediate regime the unflavoured 
scenario may correspond to the ``freeze-out" 
type leptogenesis, while the flavoured 
one -- to the ``freeze-in" type.
The correct sign can always be recovered by switching 
$\omega$ from $\pi/4$ to $3\pi/4$, or vice versa.
}. 

We find that the enhancement of the baryon asymmetry due to flavour 
effects depends strongly on the CPV phase $\delta$.
This is illustrated in Fig. \ref{fig:VIA20GeV}, 
which shows a ``freeze-in'' scenario of leptogenesis  
for $\delta = 300^\circ$, in which, in contrast to 
case with $\delta = 0$ reported in  Fig. \ref{fig:VIA15GeV},
the flavour enhancement of the baryon asymmetry is 
approximately by a factor of 60. 
The results presented in Fig. \ref{fig:VIA20GeV} 
show also that, depending on the values of 
leptogenesis parameters, 
all three lepton CPV asymmetries  
residing in the electron, muon and tauon flavours
can give significant contributions to the 
baryon asymmetry so that  
$\eta_B = \eta_{Be} + \eta_{B\mu} + \eta_{B\tau}$.
  
 The features reported in the preceding discussion
can be obtained for other choices of the parameters 
and we find, in general, that in the mass range of interest, 
varying  $x^{(0)}$ and $\delta$ accordingly, flavour effects 
can lead to enhancement 
of the generated baryon asymmetry
by a factor ranging from a few to a few hundred.

In Fig. \ref{fig:VIA16GeV} instead, the final baryon asymmetry 
is generated after all the three lepton flavour CPV asymmetries, initially 
generated during the production of RH neutrinos, are fully erased 
by the washout processes. Therefore this corresponds to the 
``freeze-out" RL case. Since the dominant 
lepton flavour related asymmetries are $\eta_{B\mu}$ and 
$\eta_{B\tau}$ 
and $\eta_B \cong \eta_{B\mu} + \eta_{B\tau}$,
the flavour effects are not 
significant
in this scenario. 
However, it is quite remarkable that in the case of the same initial 
condition -- zero initial abundance of $N_i$, the ``freeze-in" mechanism 
of baryon asymmetry generation can transform into ``freeze-out" mechanism 
for different choices of the parameters.

 In Fig. \ref{fig:xiMassVIA}, we show in the $\xi-M$ plane the region of 
successful RL in the VIA case. With respect to the TIA case, 
the region of successful RL extends to lower masses. 
In particular, we find that the minimal lower bound on the mass is reached for  
for $x^{(0)}\approx 10^6$ and reads $M\approx 0.3$ GeV.

It follows from Fig. \ref{fig:xiMassVIA}, in particular,
that the maximal value of  $\xi$ in the VIA case 
is $\xi\cong 3.3$ and is obtained, like in the TIA case,
for $M\cong 100$ GeV and $x^{(0)} \cong 1-10$. 
In the case of $M\cong 1 $ GeV we get ${\rm max}(\xi) \cong 2.5$.
This corresponds to 
${\rm max}(\sum_{l, i}|(RV)_{li}|^2)\cong 4.4\times 10^{-9}$.

As in the TIA case, we do not find any significant dependence of 
the results shown in Fig. \ref{fig:xiMassVIA} on $\alpha_{23}$ when 
$\alpha_{23}$ is varied in the interval $(0,4\pi]$. More specifically,\\ 
i) the lower bound on $M$ is always at 
$M \simeq 0.3$ GeV and is reached for $x^{(0)} = 10^{6}$;\\
ii) ${\rm max}(\xi) \simeq 3.3$ takes place at $M = 100$ GeV and 
$x^{(0)} = 1-10$ for any choice of $\alpha_{23}$;\\
iii) at $M = 1$ GeV, ${\rm max}(\xi)\simeq 2.5$ 
practically for any $\alpha_{23}$;\\
iv) at $M = 20$ GeV, ${\rm max}(\xi)\simeq 2.55~(2.63)$   
for $\alpha_{23} = 0\,(2\pi)$.\\
We notice that the curves corresponding to fixed values of  
$x^{(0) }\gtrsim 10^{4}$ show some dependence on $\alpha_{23}$. 
However, this has a minor effect on the full white region in Fig. 
 \ref{fig:xiMassVIA}, corresponding to values of the 
parameters for which we have successful leptogenesis.

 The values of the observables 
${\rm max} (\sum_{l, i}|(RV)_{li}|^2)$  and 
${\rm min} (\sum_{l, i}|(RV)_{li}|^2)$
for $M = 10$, 20, 30, 40, 50, 60, 70, 80, 85 GeV
we find in the VIA case are the same as those reported for the 
TIA case at the end of Section \ref{sec:TIA}.
This is in agreement with the fact that in the 
freeze-out leptogenesis we are analysing,  
in which the observed BAU is generated 
in the strong wash-out regime, there 
is no dependence on the initial conditions.

\vspace{0.3cm}
We emphasise that our analysis does not include the 
$N_1 - N_2$ oscillation mechanism proposed in 
\cite{Akhmedov:1998qx,Asaka:2005pn}, 
and thus our results rely purely on $1 \leftrightarrow 2$ decay 
and $2 \leftrightarrow 2$ scattering  processes in the RL case.
\begin{figure}[t]
\centering
\includegraphics[width=\textwidth]{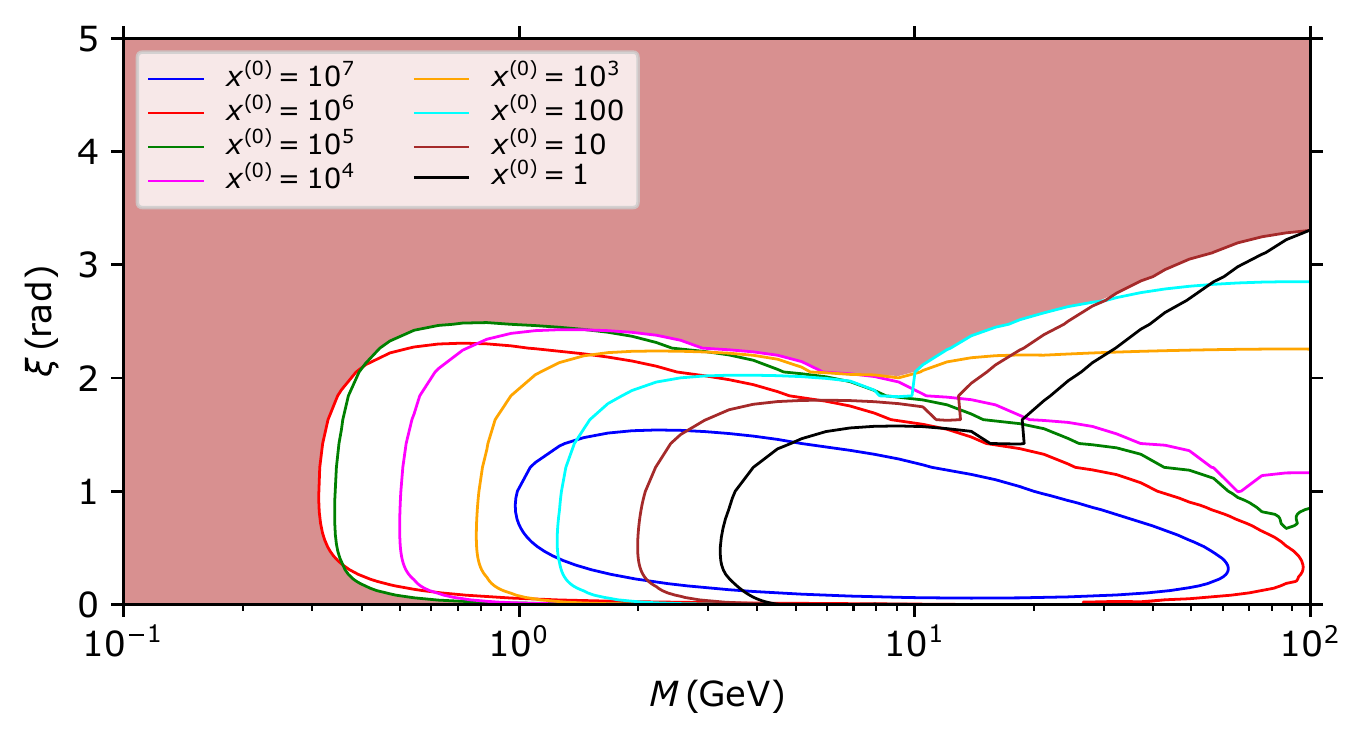}
\caption{
The allowed region (white) in the $\xi$-M plane for which 
we can always have successful resonant leptogenesis
by varying $x^{(0)}$, $\omega$ and/or the CPV phases in the PMNS matrix,
in the case of VIA.
For values in the red region the baryon asymmetry is always too small 
compared to that observed today. 
The solid contours corresponds to the maximal value of $\xi$
and $x^{(0)} = 10^7$ (blue), $10^6$ (red),
 $10^5$ (green)
$10^4$ (magenta), $10^3$ (orange), $10^2$ (cyan), 
10 (brown), and 1 (black),
for which the predicted asymmetry coincides with the observed one. 
The figure is obtained for $\delta = 3\pi/2$. 
See text for further details.
}

\label{fig:xiMassVIA}
\end{figure}
%

%
\section{Conclusions}
\label{sec:concs}
%
%
In the present article we have considered 
sub-TeV scale flavoured resonant leptogenesis (RL) 
within the minimal type-I seesaw 
scenario with two (RH) singlet 
neutrinos $N_{1,2}$ forming a pseudo-Dirac pair. 
We concentrated on the case when the masses of 
the pseudo-Dirac pair have values $M_{1,2} \ltap 100$ GeV,
$(M_2 - M_1)\equiv \Delta M \ll M_{1,2}$,
and have considered temperatures in the interval $T\sim (100 - 1000)$ GeV.
The change of the  baryon asymmetry $\eta_B$ during the generation process 
``freezes'' at the sphaleron decoupling temperature 
$T_{sph} = 131.7$ GeV and the value of $\eta_B$ at this temperature 
should be compared with the observed one.
We have not analysed in this study the scenario 
\cite{Akhmedov:1998qx,Asaka:2005pn} in which the BAU in leptogenesis
is generated via $N_1 \leftrightarrow N_2$ oscillations,
which has been extensively studied by many authors.

 We have investigated and presented results for two possible scenarios 
by which the BAU can be produced. They correspond to two different 
``initial conditions'', i.e.,  $N_{1,2}$ initial abundances 
at $T_0 >> T_{sph}$: 
i) $N_{1,2}$ thermal initial abundance (TIA), and 
ii) $N_{1,2}$  vanishing (zero) initial abundance (VIA).
In our analyses we took into account both the relevant 
$1\leftrightarrow 2$ decays and inverse decays
and $2\leftrightarrow 2$ scattering processes
including the thermal effects. 

 In the case of TIA the baryon asymmetry is produced via the so-called 
``freeze-out" mechanism, i.e., by out-of-equilibrium
processes when $N_{1,2}$ essentially coincides with their respective thermal 
abundances (Figs. (\ref{fig:TIA10GeV}) 
and (\ref{fig:xiMassTIA})). We find that in this case 
successful RL is possible for $M_{1.2}$ as low 
as 5 GeV, $M\gtap 5$ GeV. 
The flavour effects are not so relevant in the generation of the 
baryon asymmetry 
leading only to a moderate enhancement approximately 
by a factor of  2-3 of the asymmetry.

Our results show that in the VIA case one can have a 
successful leptogenesis for $M_{1,2}$ lying in the interval 
$M_{1,2} \cong (0.3 - 100)$ GeV 
(Figs. \ref{fig:VIA15GeV}, \ref{fig:VIA20GeV}, \ref{fig:VIA16GeV}
and \ref{fig:xiMassVIA}).
For the values of $M_{1,2}$ in this interval 
and depending on the values of the other leptogenesis parameters,
the baryon asymmetry can be generated 
i) either during the production of the heavy Majorana neutrinos 
$N_{1,2}$ corresponding to the ``freeze-in'' leptogenesis scenario 
(Figs. \ref{fig:VIA15GeV} and \ref{fig:VIA20GeV}),
or ii) after the produced $N_{1,2}$ abundance reached the 
thermal equilibrium value, corresponding to the ``freeze-out'' 
leptogenesis scenario (Fig. \ref{fig:VIA16GeV}). 
It is quite remarkable that in the case of the same 
initial condition -- vanishing (zero) initial abundance of $N_i$ (VIA), 
one can have a successful leptogenesis by both 
the ``freeze-in'' and the ``freeze-out'' mechanisms, 
each of the two scenarios being operative in different regions 
of the relevant parameter space.
 
 We have shown also that in the VIA case 
the flavour effects, in general, play a particularly 
important role for having successful leptogenesis. 
In certain cases they can lead to an enhancement of the baryon 
asymmetry by a factor of a few hundred  
(Fig. \ref{fig:VIA15GeV}) with respect to 
asymmetry produced by neglecting the flavour effects, i.e., 
within the unflavoured leptogenesis scenario. 
The magnitude of the flavour enhancement exhibits strong dependence 
on the value of the Dirac CPV phase $\delta$.

 Our results show further that,  
depending on the values of the leptogenesis parameters, 
the dominant among the three 
flavour components of the baryon asymmetry,
$\eta_{B_e}$, $\eta_{B_\mu}$ and $\eta_{B_\tau}$,
generated by the  corresponding  
CP violating (CPV) $e$-, $\mu$- and 
$\tau$- flavour (lepton charge)
asymmetries, could be the $\eta_{B_e}$, or
the sum  $\eta_{B_\mu} + \eta_{B_\tau}$ 
or else the contribution from all three 
components can be significant.  
Thus, the total baryon asymmetry, 
$\eta_B$ can originate either
from the CPV  $e$-flavour
asymmetry, $\eta_B \cong \eta_{B_e}$,
or from the  CPV $\mu$- and $\tau$-flavour 
asymmetries, 
$\eta_B \cong \eta_{B_\mu} + \eta_{B_\tau}$,
or else from all three 
the CPV flavour asymmetries,
$\eta_B = \eta_{B_e} + \eta_{B_\mu} + \eta_{B_\tau}$.

To summarise, we have shown that 
RL can be successful across the whole of the experimentally 
accessible region of $M_{1,2} \cong (0.3 - 100)$ GeV. 
Furthermore, we have found that leptogenesis at 
the considered sub 100 GeV  scales is compatible with 
values of the charged and neutral current 
couplings of $N_{1,2}$ in the weak interaction Lagrangian,
whose squares for, e.g.,  $M_{12} = (10 - 85)$ GeV
are in the range of $3\times 10^{-10} - 10^{-12}$.
This may pose challenges for testing the considered 
leptogenesis scenario in low-energy experiments.
A large part, if not all, of the indicated range, 
however, can be probed in the experiments at 
the discussed future FCC-ee facility 
\cite{EuropeanStrategyforParticlePhysicsPreparatoryGroup:2019qin,Blondel:2021ema}.

\section*{Acknowledgements}

S.T.P. would like to thank M. Quiros for very useful correspondence. 
We thank J. Klari\'c and  I. Timiryasov for clarifying comments 
concerning the article \cite{Klaric:2020lov}.
This work was supported in part by the INFN program on Theoretical 
Astroparticle Physics (A.G. and S.T.P.) and by the  World Premier 
International Research Center   Initiative (WPI Initiative, MEXT), 
Japan (S.T.P.). K.M. acknowledges the (partial) support from 
the European Research Council under the European Union Seventh 
Framework Programme (FP/2007-2013) / ERC Grant NuMass agreement n. [617143].

\bibliography{FRLatsubTeV}{}
\bibliographystyle{JHEP}

\end{document}